\documentclass[twocolumn,usenames,dvipsnames]{aastex631}

\submitjournal{ApJ}

\usepackage{comment, makecell,array,multirow, chngcntr}
\usepackage{xcolor}
\usepackage{lineno}
\usepackage{graphicx}

\begin{document}

\author[0000-0001-8764-7832]{J{\'o}zsef Vink{\'o}}
\affiliation{Department of Astronomy, University of Texas at Austin, 2515 Speedway, Stop C1400
Austin, Texas 78712-1205, USA}
\affiliation{Konkoly Observatory,  CSFK, Konkoly-Thege M. \'ut 15-17, Budapest, 1121, Hungary}
\affiliation{ELTE E\"otv\"os Lor\'and University, Institute of Physics, P\'azm\'any P\'eter s\'et\'any 1/A, Budapest, 1117 Hungary}
\affiliation{Institute of Physics, University of Szeged, D\'om t\'er 9, Szeged, 6720, Hungary}

\author[0000-0002-0977-1974]{Benjamin P. Thomas}
\affiliation{Department of Astronomy, University of Texas at Austin, 2515 Speedway, Stop C1400
Austin, Texas 78712-1205, USA}

\author[0000-0003-1349-6538]{J.\ Craig Wheeler}
\affiliation{Department of Astronomy, University of Texas at Austin, 2515 Speedway, Stop C1400
Austin, Texas 78712-1205, USA}

\author[0000-0002-9017-3567]{Anna Y. Q.~Ho}
\affiliation{Department of Astronomy, Cornell University, Ithaca, NY 14853, USA}

\author[0000-0002-2307-0146]{Erin Mentuch Cooper}
\affiliation{Department of Astronomy, University of Texas at Austin, 2515 Speedway, Stop C1400 Austin, Texas 78712-1205, USA}

\author[0000-0002-8433-8185]{Karl Gebhardt}
\affiliation{Department of Astronomy, University of Texas at Austin, 2515 Speedway, Stop C1400 Austin, Texas 78712-1205, USA}

\author[0000-0002-1328-0211]{Robin Ciardullo} \affiliation{Department of Astronomy \& Astrophysics, The Pennsylvania State University, University Park, PA 16802, USA} 
\affiliation{Institute for Gravitation and the Cosmos, The Pennsylvania State University, University Park, PA 16802, USA}

\author[0000-0003-2575-0652]{Daniel J. Farrow}
\affiliation{University Observatory, Fakult\"at f\"ur Physik, Ludwig-Maximilians University Munich, Scheinerstrasse 1, 81679 Munich, Germany}
\affiliation{Max-Planck Institut f\"ur extraterrestrische Physik, Giessenbachstrasse 1, 85748 Garching, Germany}

\author[0000-0001-6717-7685]{Gary J. Hill}
\affiliation{McDonald Observatory, The University of Texas at Austin, 2515 Speedway, Stop 1402, Austin, TX 78712, USA}
\affiliation{Department of Astronomy, University of Texas at Austin, 2515 Speedway, Stop C1400 Austin, Texas 78712-1205, USA}

\author{Zolt\'an J\"ager}
\affiliation{Baja Observatory, University of Szeged, Szegedi \'ut POB 766, Baja, 7900, Hungary }
\affiliation{ELTE E\"otv\"os Lor\'and University, Gothard Astrophysical Observatory, Szent Imre h. u. 112, Szombathely, 9700, Hungary}
\affiliation{MTA-ELTE Exoplanet Research Group, Szent Imre h. u. 112, Szombathely, 9700, Hungary}

\author[0000-0002-0417-1494]{Wolfram Kollatschny}
\affiliation{Institut f\"{u}r Astrophysik, Universit\"{a}t G\"{o}ttingen, Friedrich-Hund Platz 1, 37077 G\"{o}ttingen, Germany}

\author[0000-0001-5561-2010]{Chenxu Liu}
\affiliation{South-Western Institute for Astronomy Research, Yunnan University, Kunming, Yunnan, 650500, People’s Republic of China}
\affiliation{Department of Astronomy, University of Texas at Austin, 2515 Speedway, Stop C1400 Austin, Texas 78712-1205, USA}

\author[0000-0002-9498-4957]{Enik{\H o} Reg{\H o}s}
\affiliation{ Konkoly Observatory,  CSFK, Konkoly-Thege M. \'ut 15-17, Budapest, 1121, Hungary}

\author[0000-0003-0926-3950]{Kriszti\'an S\'arneczky}
\affiliation{ Konkoly Observatory,  CSFK, Konkoly-Thege M. \'ut 15-17, Budapest, 1121, Hungary}

\shorttitle{SNe in HETDEX}
\shortauthors{Vink\'o et al.}

\correspondingauthor{J\'ozsef Vink\'o}
\email{vinko@konkoly.hu}


\title{Searching for Supernovae in HETDEX Data Release 3}

\begin{abstract}
We have extracted 636 spectra taken at the positions of 583 transient sources from the third Data Release of the Hobby-Eberly Telescope Dark Energy eXperiment (HETDEX\null). The transients were discovered by the Zwicky Transient Facility (ZTF) during 2018 - 2022. The HETDEX spectra have potential to classify a large number of objects found by photometric surveys for free.  We attempt to explore and classify the spectra by utilizing machine learning (ML) and template matching techniques. We have identified one transient source, ZTF20aatpoos = AT~2020fiz, as a likely supernova candidate, and another object, ZTF19abdkelq, as a possible supernova.  We classify AT~2020fiz as a Type IIP supernova observed $\sim 10$ days after explosion, and we propose ZTF19abdkelq as a likely Type Ia SN caught $\sim 40$ days after maximum light. ZTF photometry of these two sources are consistent with their classification as supernovae. Beside these two objects, we have confirmed several ZTF transients as variable AGNs based on their spectral appearance, and determined the host galaxy types of several other ZTF transients. 
\end{abstract}


\section{Introduction} \label{sec:intro}

The advent of deep, all-sky, untargeted surveys, such as the Zwicky Transient Facility (ZTF, \citealt{bellm19}), the Dark Energy Survey (DES, \citealt{bernstein12, kessler15})  and the upcoming Legacy Survey in Space and Time (LSST, \citealt{ivezic19}) has ushered in a new age for the study of transient astronomy, and it is now one of the most popular fields in astronomy and astrophysics. The thousands of new transients discovered each day offer the potential to build statistical samples of otherwise rare and poorly known phenomena, such as tidal disruption events \citep{gezari21} and superluminous supernovae \citep{galyam19}, as well as find new types of transient objects that currently remain unknown. 

Supernovae (SNe) are among the most well-known and frequently studied transient objects for several reasons. First, due to their outstanding peak brightness, they can be discovered and studied even at high redshifts, i.e., $z > 1$. Moreover, the $\sim$month timescale of their evolution is neither too fast nor too slow compared to human timescales, which makes them ideal targets for astrophysical studies. Also, supernovae are among the best distance indicators in extragalactic astronomy, and thus, are useful from not only an astrophysical, but also a cosmological point of view:  for example, the discovery of the accelerated expansion of the Universe was at first based on distance measurements from Type Ia SNe \citep{riess18, perlmutter99}. Finally, the James Webb Space Telescope (JWST) and future space observatories will extend our cosmic horizon considerably, thus, enabling the discovery of transients beyond $z > 2$, in the very early Universe \citep[e.g.][]{rv19,rvz20,rvs21}.
The recently discovered very high redshift galaxies at $z > 12$ -- including one at $z = 16.7$ -- by JWST
\citep{donnan22} improve our understanding of the cosmic star formation rate at $z > 8$ and the epoch of reionization. Studies of these high redshift supernovae will ultimately constrain the (true) initial mass function of the first stars. 

The unprecedented number of new transients in the data stream makes their analysis very challenging. Since most of the surveys mentioned above are based on imaging and photometry, the possibility for a prompt revealing of the physical nature of new transients is limited. As most classification schemes require (mostly optical) spectroscopy, it is not currently possible to definitively classify the bulk of these new sources. Instead, the current strategy is concentrated on identifying potentially interesting sources based on either their position (e.g., being in a nearby galaxy or galaxy cluster) or their photometric properties (e.g., unusually bright, blue/red, fast/slow evolution, etc.). This still leaves most transients unclassified and they fade away with minimal spectroscopic attention. 

In such circumstances, survey programs that produce a massive number of spectroscopic observations can be useful. One of them is the 
Hobby-Eberly Telescope Dark Energy eXperiement (HETDEX, \citealt{gebhardt21,hill21}). HETDEX is a blind spectroscopic survey that, once completed, will use observations of 1.2 million Ly$\alpha$ emitting galaxies to measure the baryonic acoustic oscillations (BAO) and constrain the fundamental parameters that describe the time evolution of the dark energy equation-of-state. HETDEX aims to obtain spectral coverage of 540 deg$^2$ of the sky with a filling factor of 1 in 4.6; a spectrum is obtained at every position that is covered by a fiber in the field-of-view, including all objects that are brighter than the HETDEX $5\sigma$ limiting magnitude of $m_{\rm lim}(\mathrm{AB}) \sim 22.5$ in $g$-band \citep{gebhardt21}.

The HETDEX survey is unique as it contains untargeted, deep, wide-field spectroscopic data.  While HETDEX is at heart a high-redshift galaxy survey \citep{davis21, lujan22}, it also obtains spectra of all other objects in its field-of-view, including stars of all types \citep{hawkins21}, early and late type galaxies \citep{indahl21}, high-redshift quasars \citep{zhang21, liu22}, and, in principle, supernovae. The unbiased nature of the HETDEX spectroscopic selection function implies that there must exist spectra of anomalous, perhaps unprecedented, objects within the HETDEX continuum-source catalog. The HETDEX data can also be used, at least potentially, for the classification of all types of transient objects that appear within the survey footprint.

Another advantage of HETDEX is that it uses integral field units (IFUs), thus, not only single objects but also their immediate environment can be observed spectroscopically. This could be promising, for example, for studying the hosts of extragalactic transients (see Section~\ref{sec:hosts}).

In this paper we present the first attempt to use HETDEX data for spectroscopic classification and characterization of transients that were discovered within the survey's footprint. While our primary purpose was finding supernovae, it turned out that the HETDEX spectra are also useful for identifying variable active galactic nuclei (AGN), and studying the host galaxies of supernovae and other transients. 

This paper is organized as follows. Data extraction is outlined in Section~\ref{sec:data}, then the applied classification methods are detailed in Section~\ref{sec:methods}. Results are discussed in Section~\ref{sec:disc}, while the conclusions are summarized in Section~\ref{sec:conc}. 

\section{Data}\label{sec:data}

HETDEX uses the Visible Integral-field Replicable Unit Spectrograph (VIRUS, \citealt{hill21}) instrument at the focal surface of the Hobby-Eberly Telescope (HET, \citealt{ramsey98,hill21}). VIRUS comprises up to 78 integral field units, each with 448 $1\farcs 5$-diameter fibers, such that 33,152 individual spectra covering 54~arcmin$^2$ are obtained with each visit. For more details about the survey design concepts and specifications, see \citet{gebhardt21}. For more details about the instrumentation, see \citet{hill21}.

The HETDEX Data Release 2.1 (HDR2.1) continuum-source catalog contains spectra for ${\sim} 62$k unique objects (Mentuch Cooper, in prep), and includes a vast amount of information that is too large to be visually classified by human experts. Thus, the very nature of the HETDEX survey design motivates machine learning efforts not only for anomaly detection, but also for object classification. In the HETDEX Data Release 3 (HDR3), the size of the continuum-source catalog has been extended further, reaching ${\sim} 230$k sources. 


Supernovae identifications in the HETDEX catalog are challenging, not least because they are not the ultimate target of the survey. As a result, the various reduction pipelines that produce the survey's data products are not optimized to find extragalactic transients. For example, one such issue is that, within the continuum detection process, sources are rejected if they are within $1\farcs 5$ of another galaxy; in these cases only the brightest continuum object remains in the catalog. This cut may discard supernovae and similar astronomical transients that are associated with a bright host galaxy, but are actually separate point sources near their host galaxy's core. We therefore opted not to use the default HETDEX data products for our initial search for supernovae. 

Instead, for this work we chose to perform forced spectral extractions at the known locations of transient sources. There are many astronomical surveys that can provide auxillary information to HETDEX\null. One of these is the Zwicky Transient Facility (ZTF, \citealt{bellm19,graham19,masci19,dekany20}), which covers the full northern sky in two filters every three nights.  This survey typically produces between $10^5$ and $10^6$ alerts per night \citep{Forster21}, 80\% of which are associated with real astronomical sources \citep{CarrascoDavis21}. 
Thus, we aimed to extract spectra at the positions of public ZTF transients announced between 2018 and 2022 from the HETDEX spectroscopic dataset, in the hopes of finding SN-like signals (broad features) in those spectra. Details are given in the rest of this paper. 

\subsection{Source extraction}\label{sec:extraction}

We used the Automatic Learning for the Rapid Classification of Events (ALeRCe; \citealt{Forster21}) Application Programming Interface (API)\footnote{https://alerce.readthedocs.io/en/latest/tutorials/ztf{\_}api.html} to query the ZTF servers to find all detected ZTF transients that were within 12 arcminutes of the position of a HETDEX pointing.  (This is roughly the size of the HET's usuable focal plane, which is 20 arcminutes in diameter.) We then determined which of the resulting 4585 ZTF sources fell within 3\farcs5 of any of the $\sim$34,000 fibers that are part of VIRUS\null. This additional downsampling resulted in 636 spectra of 583 objects matched between ZTF and HETDEX\null. (Although HETDEX is a single epoch survey, some science verification fields have been observed multiple times.)
Using \texttt{HETDEX-API/get\_spec.py}\footnote{\url{https://github.com/HETDEX/hetdex_api/blob/master/hetdex_tools/get_spec.py}}, we were then able to extract HETDEX spectra that are closely matched to each of these transient locations. We used a point spread function (PSF)-weighted 1D extraction at the exact position that the ZTF source was detected and required that at least 7 fibers be contained within the 3\farcs5 circular aperture that is needed to fully capture the VIRUS PSF\null.  
Thus, the final spectrum is produced from a PSF-weighted sum over many fibers combined within a circular aperture with a radius of 3\farcs5 \citep{gebhardt21}.

It is worth noting here that the HETDEX spectra associated with these sources may not have been taken while the prospective transient was active. Indeed, in the majority of cases they were not. We can, however, still search for the nebular spectra of supernovae that have faded beyond the ZTF detection limit. While searches for pre-SN progenitor stars may also be possible, it is likely that those progenitors will be too faint for detection.


\section{Methods} \label{sec:methods}

This section summarizes the various methods that we applied to explore and characterize the extracted HETDEX spectra.

\subsection{Exploration: machine learning (ML)}
\label{sec:ml}

A large number of discoveries in astronomy, and science more generally, have been catalyzed due to some degree of serendipity. The use of machine learning for anomaly detection has the potential to automate serendipity, and dramatically widen the potential discovery pool.

One method that facilitates the visual exploration of large datasets is t-distributed stochastic neighbor embedding (t-SNE; \citealt{vanderMaaten08}). The t-SNE program is fundamentally a dimensionality-reduction technique that preserves the pairwise similarity between points while producing a $d$-dimensional map of those points (usually $d=2$). We use the version of t-SNE implemented in the {\tt scikit-learn} Python package \citep{scikitlearn11}. 

Here the dimensionality of the high-order space is set by the number of flux measurements (1036) in a single HETDEX spectrum. The t-SNE program works by first computing pairwise similarities $p_{ij}$ between these high-dimensional data points ${\bf x}_i$ and ${\bf x}_j$,
\begin{equation}
    p_{j|i} = \frac{{\rm exp}(-||{\bf x}_i - {\bf x}_j||^2/2\sigma^2_i)}{\sum_{k \neq i}{\rm exp}(-||{\bf x}_i - {\bf x}_k||^2/2\sigma^2_i)},
\end{equation}
where $\sigma_i$ is the variance of a Gaussian distribution centered on point ${\bf x}_i$. This parameter is related to a user-defined hyperparameter called the perplexity. We choose a perplexity value of $\sqrt{N}$, where $N$ is our number of data points. 

To mitigate problems associated with outliers, the symmetrical pairwise similarity is defined as:
\begin{equation}
    p_{ij} = \frac{p_{j|i} + p_{i|j}}{2N},
\end{equation}
where $p_{j|i}$ is the probability that point ${\bf x}_{i}$ would choose point ${\bf x}_j$ as its nearest neighbor under the assumption of a Gaussian probability distribution centered at point ${\bf x}_i$.

The next step is for t-SNE to attempt to learn a low-dimensional ($d=2$) representation of the points, ${\bf y}_1,...,{\bf y}_N$, that preserves the high-dimensional similarities $p_{ij}$ as analogous low-dimensional similarities $q_{ij}$. In this low dimensional map, 
\begin{equation}
    q_{ij} = \frac{(1 + ||{\bf y}_i - {\bf y}_j||^2)^{-1}}{\displaystyle\sum_{k \neq m}(1 + ||{\bf y}_k - {\bf y}_m||^2)^{-1}} .
\end{equation}
This metric is related to the Student's t-distribution \citep[e.g.,][]{numrec}, and its primary purpose is to avoid over-crowding in the low-dimensional embedded space. Also, the Student's t-distribution is faster to compute compared to a Gaussian as it does not contain any exponential terms.

The key process of t-SNE is to determine the low-dimensional embeddings of the points ${\bf y}_i$ by minimizing (via gradient-descent) the Kullback-Leibler divergence between the low- and high-dimensional similarity distributions $P$ and $Q$,
\begin{equation}
    KL(P||Q) = \sum_{i \neq j} p_{ij} {\rm log} \frac{p_{ij}}{q_{ij}} .
\end{equation}
This will result in an optimized $d=2$ dimensional map that preserves the similarities of points from the high-dimensional distribution.

\subsection{Classification: template matching via $\chi^2$-minimization}
\label{sec:chi2template}

The traditional method for automated classification of astronomical objects is through comparisons with a list of template spectra using a mathematical criterion to judge the quality of matching; the classification then follows from the best-matched template. We employed this template-matching scheme using as templates the Pickles Stellar Spectral Flux Library \citep{pickles98} for stars and the Kinney synthetic spectra \citep{kinney96} for galaxies. The former contains 131 stellar spectra covering the Harvard classes from O5 through M4 and the Morgan-Keenan luminosity classes from I to V; the latter includes spectra for 11 galaxies, including an elliptical (E), a lenticular (S0), three spirals (Sa, Sb, Sc) and six star-bursting (SB1 - SB6) systems. The star-forming templates are dominated by emission features of varying strength, while the elliptical and lenticular templates represent the spectra of most passive galaxies.  

We adopted $\chi^2$ statistics for measuring the quality of the match between the observed spectrum and a template. The normalized $\chi^2$ function was defined as
\begin{equation}
    \chi_N^2 ~=~ \frac{1}{N} \sum_{k=0}^{N} \frac{1}{\sigma_k^2} ( f_c^{\rm obs}(\lambda_k) - f_c^{\rm templ}(\lambda_k))^2, 
\label{eq:chi2}
\end{equation}
where $N$ is the number of sampled wavelengths in an observed spectrum, the functions $f_c^{\rm obs}$ and $f_c^{\rm templ}$ are the continuum-normalized flux of the object and the template as a function of wavelength ($\lambda$), and $\sigma_k$ is the uncertainty of the observed fluxes. Note that, as written, Equation~\ref{eq:chi2} does not contain any terms for the the possible redshift mismatch between the observed and the template spectrum, and thus is only applicable for Milky Way objects with $z=0$. Before using the equation for galaxies, the templates were therefore redshifted, and a value of $\chi^2_N$ was computed for each $z$ between $0 < z < 0.3$ range, using $dz = 0.001$ (see Section~\ref{sec:disc-templ} for further information).  

Before applying Equation~\ref{eq:chi2},
simple linear interpolation was applied to the template spectra to correct for the different wavelength sampling between the template and the observed spectrum.  In addition, both spectra were normalized to the continuum by iteratively fitting an 8th order polynomial to both the observed and the template spectrum. After each iteration, the standard deviation ({\tt stddev}) of the residuals were calculated and data that exceeded the residual by {\tt -2*stddev} and {\tt +3*stddev} were rejected from the sample. After $\sim 10$ such iterations the fitted polynomial was found to represent the continuum reasonably well. 

The uncertainties of the observed fluxes were estimated empirically by calculating the mean ($\langle f_c \rangle$) and standard deviation ($\sigma$) of the continuum-normalized data in the red part of the spectrum, between 4600 and 5400\,\AA\null. The same region was used to derive an empirical signal-to-noise (S/N) parameter, defined as {\tt s/n}$= \langle f_c \rangle / \sigma$. 
Because of the presence of spectral features in the fitted wavelength region, this method clearly overestimates the noise in the case of spectra with moderate to high S/N, while giving more realistic results for noise-dominated data. We chose this simple approach of adopting a single $\sigma$ for the whole spectrum in Eq.~\ref{eq:chi2}, i.e., $\sigma_k = \sigma$ for all $k$, because the majority of our sources did not reach high S/N, and the $\chi^2_N$ function was used only for measuring the quality of fitting between the observed and the template spectra.  We do not treat the value  as a proper statistical quantity.

\subsection{Classification: template matching via cross-correlation}
\label{sec:sntemplate}

\begin{figure*}
    \centering
    \includegraphics[width=8cm]{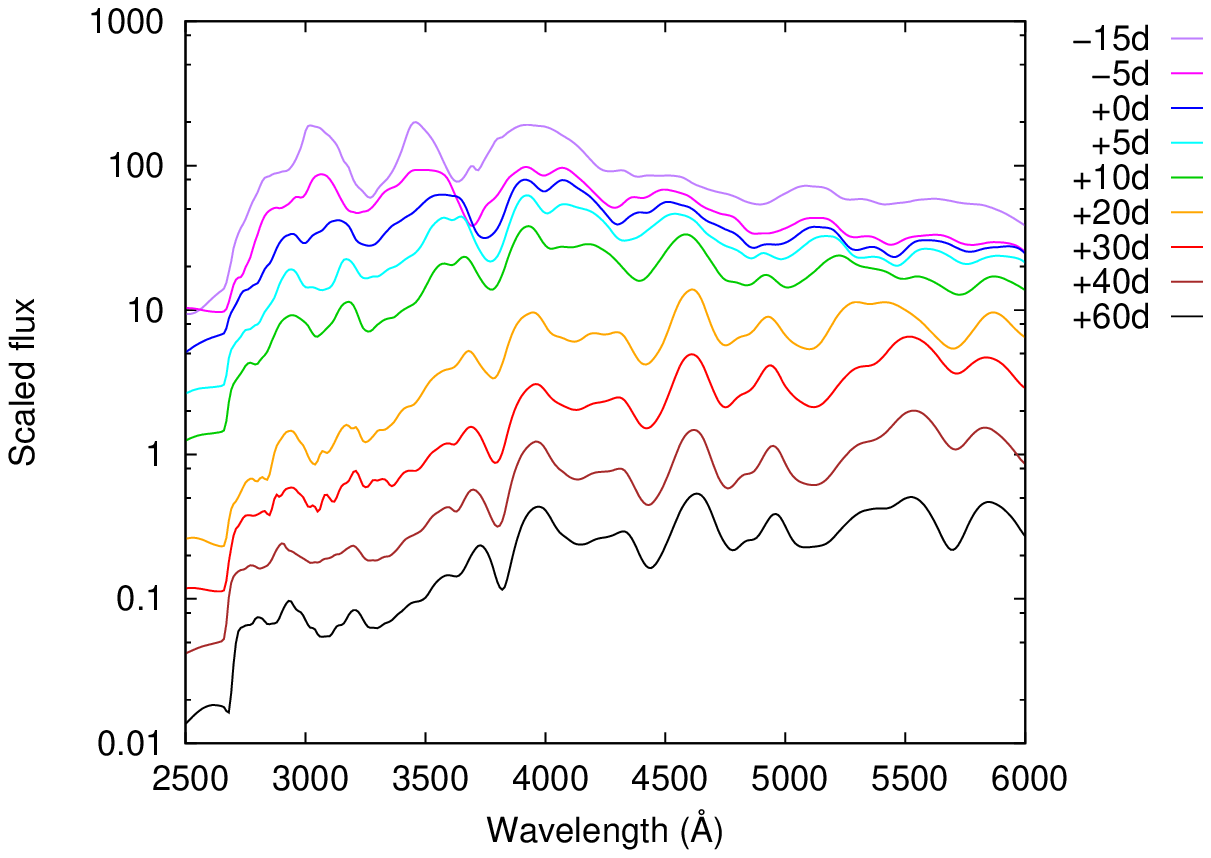}
    \includegraphics[width=8cm]{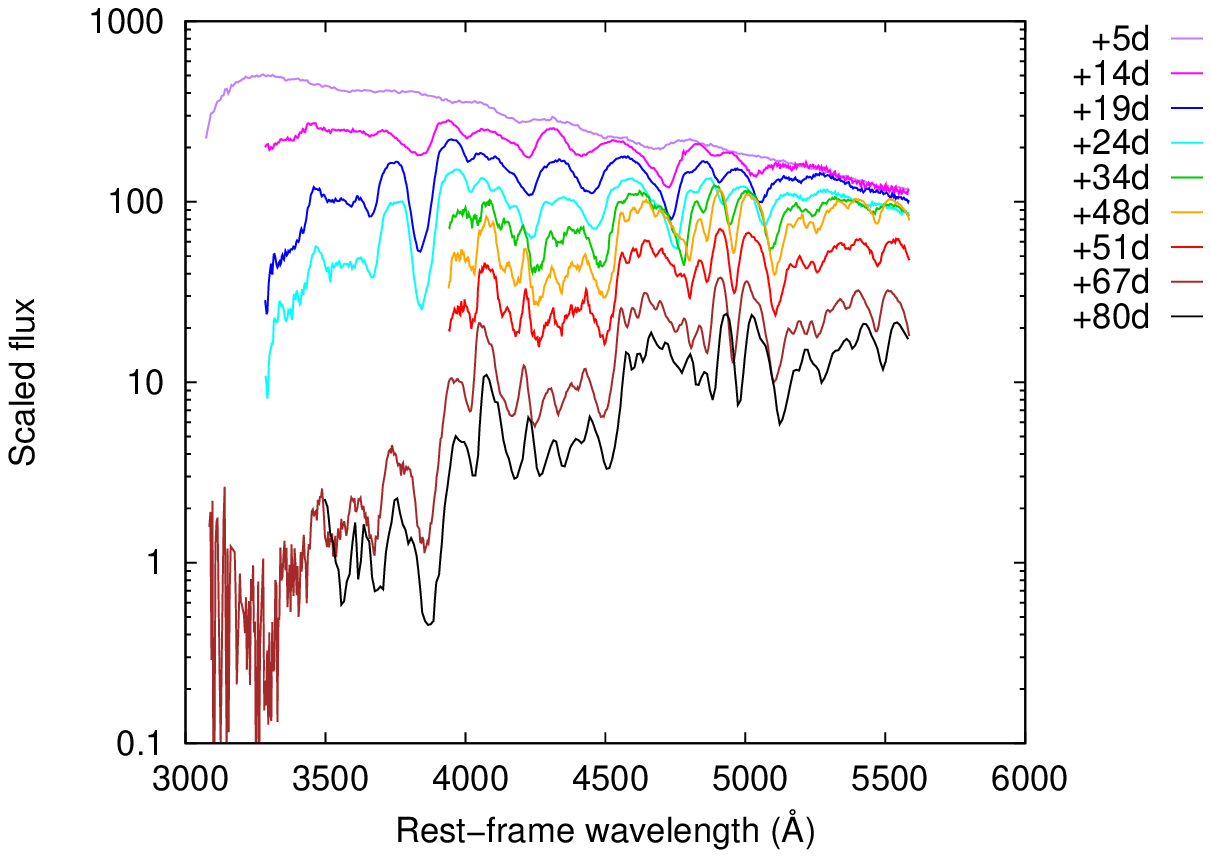}
    \caption{The series of supernova spectral templates for SNe Ia (left) and SNe IIP (right) applied in the \textit{CCF}-method. The spectral phase in days for each template is indicated in the legend.}
    \label{fig:sn_templates}
\end{figure*}

The other widely used algorithm for template matching is cross-correlation. A code that is based on cross-correlation and applied very often in the supernova community is {\tt SNID} \citep[SuperNova IDentification;][]{snid}, which uses the observed spectra of different types of SNe as templates. Such codes compute the cross-correlation function (\textit{CCF}) between the observed ($f(\lambda)$) and the template spectrum ($g(\lambda$)) via the following definition:
\begin{equation}
    CCF(\Delta x) = \int_{-\infty}^{+\infty} f(x) \cdot g(x - \Delta x) dx, 
\end{equation}
where $x = \ln \lambda$, $\Delta x = \ln(\lambda / \lambda_0) = \ln(1+z)$ and $z$ is the redshift of the object. In short, the \textit{CCF} measures the overlap, i.e., the similarity, between the object and the template spectrum: the stronger the overlap, the greater than similarity of the two spectra. Thus, the height of the $CCF$ peak is proportional to the overlap, while the horizontal shift ($\Delta x$) gives the redshift of the observed spectrum with respect to the template.

Because the public template libraries of {\tt SNID} are based on published observed spectra, Type Ia SNe are overrepresented in them. In addition, most of these template libraries do not fully cover the HETDEX wavelength range of 3500 to 5500\,\AA\null. Thus, instead of applying {\tt SNID} for our sample, we developed our own pipeline based on the {\tt fxcor} task in {\tt IRAF}\footnote{IRAF was distributed by the National Optical Astronomy Observatory, which is operated by the Association of Universities for Research in Astronomy (AURA) under a cooperative agreement with the National Science Foundation.}. 

Since our primary intention was finding supernovae, we adopted the Type Ia supernova templates compiled by Eric Hsiao \citep{hsiao} for our analysis. These spectra span from $-15$ days to $+91$ days in phase with respect to the moment of $B$-band maximum, with a 1-day cadence. Since our single epoch data do not allow precise temporal resolution, we pre-selected 9 spectra from the library (having epochs at $-15$, $-5$, 0, +5, +10, +20, +30, +40, and +60 days with respect to maximum light) and used only those data in the cross-correlation. The selected SN Ia templates are shown in the left panel of Figure~\ref{fig:sn_templates}. Each template spectrum has $\Delta \lambda = 10$\,\AA\ resolution (which is much lower than the $\sim 2$\,\AA\ resolution of the HETDEX spectra) and covers the wavelength interval between 2500 and 6000\,\AA. 

Because the spectra of hydrogen-poor (Type Ib/c) core collapse SNe are often similar to those of Type Ia SNe, at least around maximum light when the SN appears to be the brightest, we did not define a separate set of templates for those objects. Instead, the Hsiao-templates were utilized for the SN Ib/c-type objects.

For hydrogen-rich (Type II) core collapse SNe, we adopted the published spectra of the archetypal Type IIP SN 1999em \citep{hamuy01, elmhamdi03}, downloaded from the Weizmann Interactive Supernova Data Repository (WiseRep)\footnote{\tt https://www.wiserep.org/}. The selected spectra are corrected for the redshift of SN~1999em ($z = 0.002392$) and shown in the right panel of Figure~\ref{fig:sn_templates}. We chose only the spectra that were taken during the plateau phase (lasting $\sim 100$ days after the explosion), because the SN is brighter by at least 1.5 - 2 magnitudes during this period than in the subsequent tail- and nebular phase.   Thus, it is more probable to detect a Type~II at this time.

\section{Results and Discussion} \label{sec:disc}

The results of our analysis are presented and discussed in this Section. 

\subsection{Machine learning}
\label{sec:disc-ml}

\begin{figure}
    \centering
    \includegraphics[width=\columnwidth]{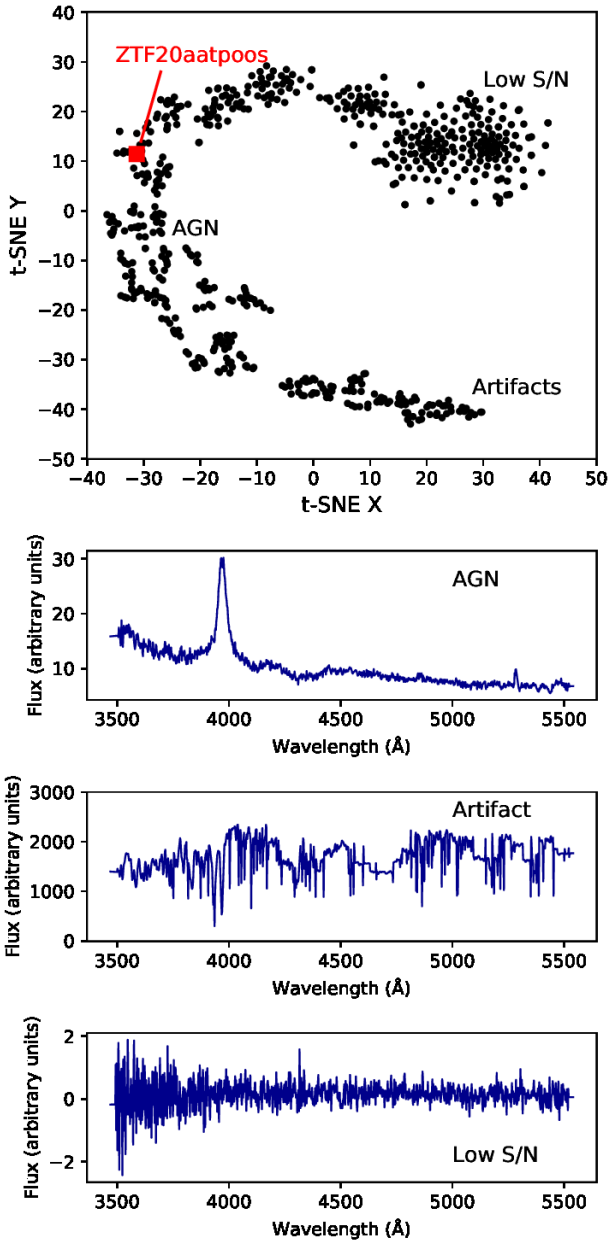}
    \caption{{\it Top panel:} The output embedding of our 636 HETDEX-ZTF spatially-matched spectra from t-SNE\null. Each point represents a HETDEX spectrum, and spectra that are similar are clustered together. The classes that dominate each region of the map are marked in the text. ZTF20aatpoos is shown as a red square, and is found in a map region that is otherwise primarily inhabited by passive galaxies. {\it Bottom panels:} Example spectra of AGN, pipeline artifacts, and low signal-to-noise objects are shown for the three categories marked on the top panel.}
    \label{fig:tsne}
\end{figure}

After having extracted spectra from the HETDEX database at the positions of the known ZTF transients (see Section~\ref{sec:extraction}), we normalized the spectra by their uncertainty (produced by the HETDEX reduction pipeline) to account for edge defects and other minor issues, and fed them directly into the t-SNE algorithm\footnote{Some other studies have used a principle component analysis to reduce the dimensionality of the data before feeding it into t-SNE. We found that this was not necessary in our case due to the relatively small size of the data set.}. This resulted in a two-dimensional map shown in Figure \ref{fig:tsne}, which we could then explore and visually inspect.  We created a point-and-click tool to quickly and efficiently examine the spectra represented by each t-SNE data point, and by this visual inspection process, we identified the clusters that were predominantly AGN, star-forming galaxies, and low S/N sources. We also flagged several objects for further analysis. These objects of special interest were analyzed using {\tt SNID} in order to determine their astronomical type. Amongst these was ZTF20aatpoos, which we classified as a Type II-P SN about 10 days after peak (see Section~\ref{sec:sne}). This classification represents our first success in searching for SNe in the HETDEX database, and motivates efforts for further discovery.

\subsection{Template fitting via $\chi^2$}
\label{sec:disc-templ}

The $\chi^2_N$ value (Equation~\ref{eq:chi2}) was computed for all 636 observed spectra using all templates in both the stellar and the galaxy library. We defined three groups of objects, named {\tt star}, {\tt galaxy} and {\tt low}, as the initial classification categories for the observed spectra.  Spectra that were noise-dominated and did not reach {\tt s/n}=3 were moved into the {\tt low} category. Sources that had lower $\chi^2_N$ values for fits to $z = 0$ stellar templates than fits to galaxies at $z>0$ were classified as stars, and the remaining spectra were, at first, labeled as galaxies.

In the next step, all spectra were inspected visually, and peculiar objects were moved into additional categories. We identified broad-lined AGNs (category {\tt agn}) in the {\tt galaxy} group using the measured width of their emission features.  Since many of these sources showed only a single broad line in the relatively narrow wavelength interval observed by HETDEX (3500 to 5500\,\AA), their redshifts, as derived from the best-fitting galaxy template may differ substantially from the true value. These objects are analyzed further in Section~\ref{sec:agn}. Finally, spectra that did not resemble any other members in the above categories were moved to the {\tt uncertain} category. 

\begin{table}
    \centering
     \caption{Statistics of classified objects by their ALeRCE categories}
    \begin{tabular}{lcccccc}
    \hline \hline
    Sp. class & total & VS & AGN & SN & AST & UNC \\
    \hline
    {\tt star} & 149  & 114 & 18 & 6 & 0 & 11  \\
    {\tt galaxy} & 122 & 2 & 61 & 50 & 0 & 9 \\
    {\tt agn} & 52 & 4 & 38 & 7 & 0 & 3 \\
    {\tt low} & 64 & 0 & 6 & 30 & 4 & 32 \\
    \hline
    \end{tabular}
    \label{tab:stat}
\end{table}

\begin{figure}
    \centering
    \includegraphics[width=8cm]{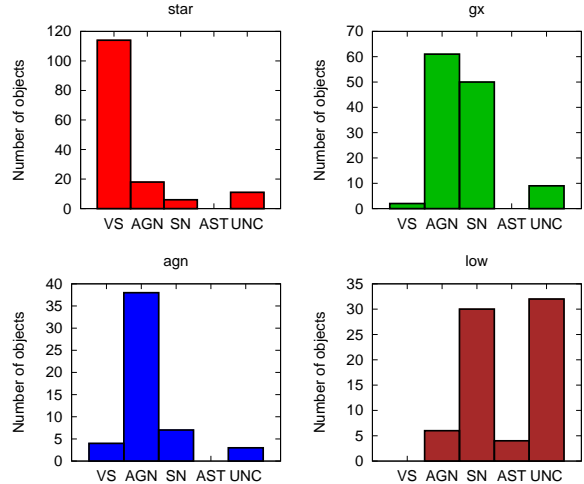}
    \caption{Sources classified via $\chi^2$-minimization as a function of their ALeRCE categories (Variable Star, AGN, SN, ASTeroid, UNClassified; see text). Each panel shows the histogram from sources belonging to the same HETDEX spectral type, namely stars (top left), galaxies (top right), AGNs (bottom left) and low S/N (bottom right panel).}
    \label{fig:classif_stats}
\end{figure}

It is an interesting question whether the results from our simple classification scheme have any correlation with the classifications performed by the ALeRCE pipeline. To test for this, we cross-compared our classification results for each transient ({\tt star, galaxy, agn,} and {\tt low}) to their most probable ALeRCE classification. From the various algorithms available in ALeRCE, we selected the one that had the highest probability value from the so-called Stamp Classification that uses the following categories: variable star ({\tt VS}), active galactic nucleus ({\tt AGN}), supernova ({\tt SN}), asteroid ({\tt AST}), and artifact ({\tt BOGUS}). In addition, we introduced a 6th category, named unclassified ({\tt UNC}), for those ZTF objects that were announced but had no ALeRCE classification available. 

As an initial step, all sources belonging to the {\tt BOGUS} category were rejected from the sample. Such ``transients'' usually appear in the vicinity of bright, saturated objects, e.g., near their diffraction spikes, and are artifacts of the ZTF image subtraction process. A total of 251 such artifacts were removed from the sample, and only the remaining 385 spectra were analyzed further. 

The summary of our cross-comparison can be found in Table~\ref{tab:stat} and is shown graphically in Figure~\ref{fig:classif_stats}. It is seen that the results from the $\chi^2$ template matching correlate very well with those from the ALeRCE Stamp Classification. For example, of the 149 sources with stellar-like HETDEX spectra, 114 were classified as variable stars by ALeRCE\null. Similarly, the 122 galaxy-like HETDEX spectra belong to ZTF transients that were classified by ALeRCE as either {\tt AGN} or {\tt SN}\null. It is especially encouraging that out of the 52 spectra that we classify as belonging to AGN, ALeRCE found 38 to be likely AGN.

As expected, many of the low S/N spectra, i.e., those that belong to the {\tt low} category in our scheme, have no ALeRCE classification. It is also interesting that most of the transients that ALeRCE identified as most likely SNe are associated with either galaxies or noise-dominated sources. This is also an expected result, because many of the HETDEX spectra were taken outside the visibility window of an active transient, resulting in either a low S/N spectrum or that of the likely host galaxy. 

\begin{figure}
    \centering
    \includegraphics[width=8cm]{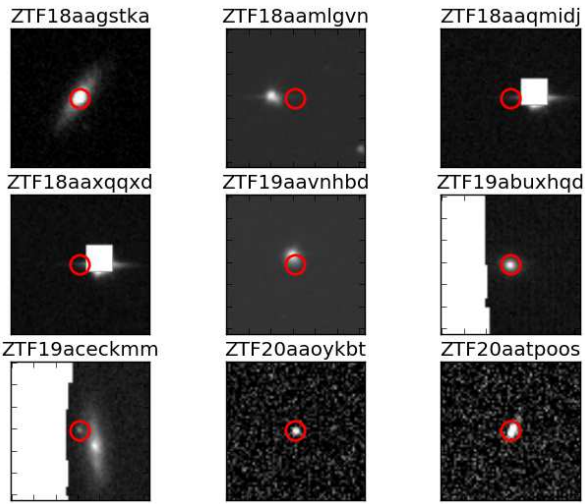}
    \caption{ZTF frame stamps (size $78"\times 78"$) taken in the $r$-band for the active ALeRCE SNe. North is up and East is to the left in each stamp. The white squares indicate masks placed onto saturated pixels.}
    \label{fig:stamps_active}
\end{figure}

\begin{figure}
    \centering
    \includegraphics[width=8cm]{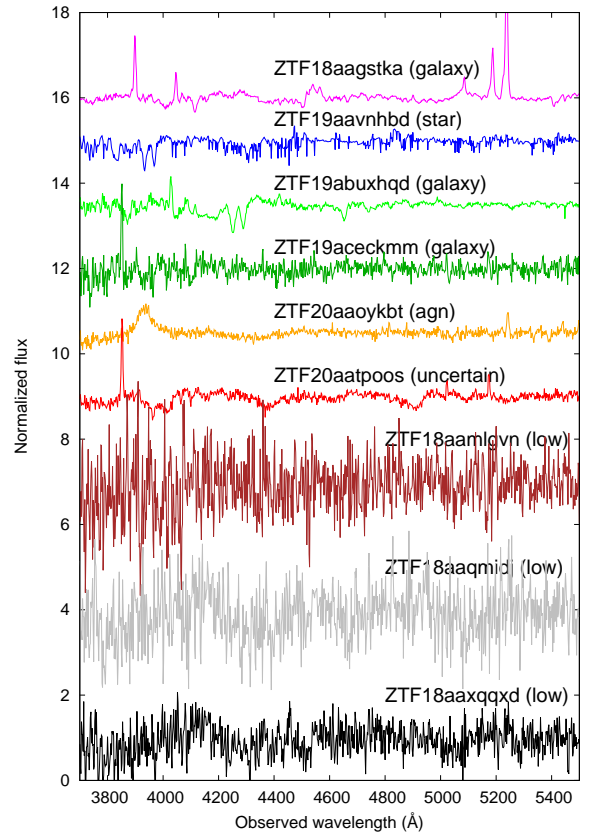}
    \caption{Normalized HETDEX spectra of the active transients classified as SNe by ALeRCE.}
    \label{fig:classif_active}
\end{figure}

\begin{table*}
\caption{Active ZTF transients classified as SNe by ALeRCE }
\begin{center}
\scriptsize
\begin{tabular}{lccccccccccl}
\hline
\hline
Name & R.A.(deg) & Dec.(deg) & MJD$_{\rm DEX}$ & $\Delta t$ (d) & AL type & AL prob. & Sp.type & S/N & $\chi^2$ & z & Class. \\ 
\hline
ZTF18aagstka & 226.937577 & 51.452847 & 58660 & 405 & SN & 0.4370 & SB4 & 3.1275 & 0.2609 & 0.045 & {\tt galaxy} \\
ZTF18aamlgvn & 230.134233 & 51.274517 & 58314 & 38 & SN & 0.4128 & -- & 1.9978 & 2.3363 & 0 & {\tt low} \\
ZTF18aaqmidj & 185.676093 & 51.375824 & 58495 & 195 & SN & 0.4317 & --  & 0.001 & 0.001 & 0 & {\tt low} \\
ZTF18aaxqqxd & 185.675409 & 51.375815 & 58495 & 223 & SN & 0.3498 & -- &  3.0660 & 2.4245 & 0 & {\tt low} \\
ZTF19aavnhbd & 267.924007 & 65.042444 & 58720 & 87 & SN &  0.4321 & G0V & 10.4723 & 1.1244 & 0 & {\tt star} \\
ZTF19abuxhqd & 15.680562 & 0.720323 & 58788 & 62 & SN & 0.3881 & S0 & 13.9679 & 3.1949 & 0.080 & {\tt galaxy} \\
ZTF19aceckmm & 172.320355 & 51.525159 & 58907 & 141 & SN & 0.7280 & SB6 & 7.4568 & 2.5238 & 0.032 & {\tt galaxy} \\
ZTF20aaoykbt & 181.108375 & 55.606322 & 58952 & -2 & SN & 0.5260 & SB6 & 11.3571 & 3.1397 & 0.056 & {\tt agn} \\
ZTF20aatpoos & 186.512456 & 55.702264 & 58954 & 18 & SN & 0.6277 & SB6 & 9.1188 & 2.1196 & 0.033 & {\tt uncertain} \\
\hline
\end{tabular}
\label{tab:active}
\end{center}
\tablecomments{\scriptsize Columns: (1) ZTF name; (2), (3): J2000 coordinates in degrees; (4): MJD of HETDEX observation; (5): Time difference between the HETDEX observation and the beginning of ZTF followup in days; (6): most probable ALeRCE type; (7): ALeRCE classification probability; (8) Spectral type of best-matching template; (9): signal-to-noise parameter; (10): normalized $\chi^2$; (11): redshift;  (12): $\chi^2$ classification. }
\end{table*}

Although our HETDEX spectral extractions were all at the announced coordinates of the ZTF transients, many of the spectra are not actually associated with a transient object, due to the object's finite visibility window.  In order to overcome this issue, we identified the ``active'' transients that were stamp-classified as {\tt SN} by ALeRCE and were observed by HETDEX when their ZTF followup was still active. Nine such objects, collected in Table~\ref{tab:active}, were found. Their ZTF subframes, downloaded from the ALeRCE website, are shown in Figure~\ref{fig:stamps_active}, while their HETDEX spectra and the results of their $\chi^2$-classifications are plotted in Figure~\ref{fig:classif_active}.  

It is seen that three of the objects that were stamp-classified as {\tt SN} by ALeRCE (ZTF18aamlgvn, ZTF18aaqmidj, and ZTF18aaxqqxd) have low S/N spectra that prevented a reliable classification.  (In fact, the latter two appear to be artifacts from the same very bright star.)  Of the remaining 6 objects, 3 are galaxies, 1 is a star and another is a broad-lined AGN\null. Of the set, ZTF19aceckmm is the most interesting, because it is located near a bright galaxy (see Figure~\ref{fig:stamps_active}), and has an ZTF light curve (shown on the ALeRCE website) that resembles that of a Type I supernova. Unfortunately, the HETDEX spectrum was taken 141 days after the ZTF discovery, when the SN had already faded below the HETDEX detection limit.   

The last object, ZTF20aatpoos, which we classified as {\tt uncertain}, is the only object in the sample that shows neither a stellar- nor galaxy-like spectrum (see Figure~\ref{fig:classif_active}). In fact, the spectrum is that of a Type IIP supernova, which was also found as a SN candidate in Section~\ref{sec:disc-ml} and in Section~\ref{sec:disc-cc} below. More details on this object are presented in Section~\ref{sec:sne}.

\subsection{Cross-correlation}
\label{sec:disc-cc}

\begin{figure*}
    \centering
    \includegraphics[width=8cm]{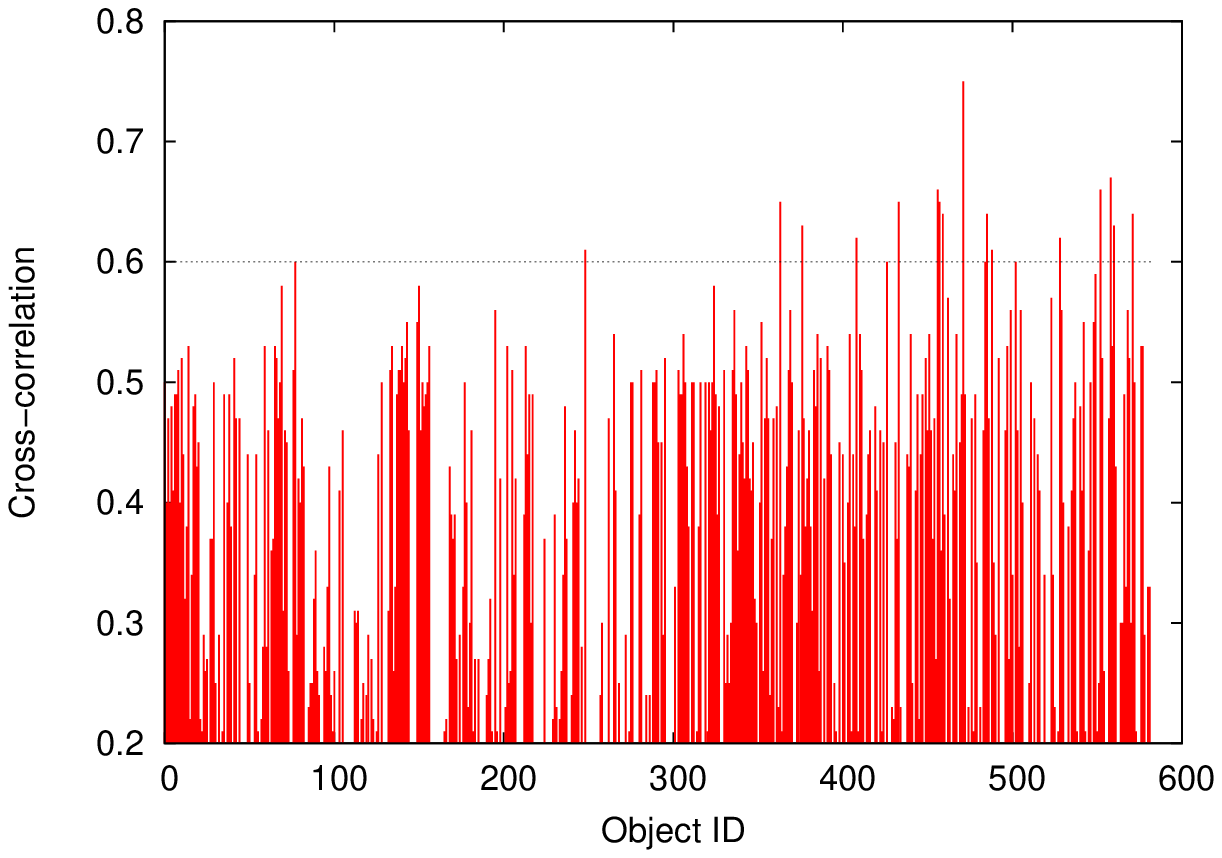}
    \includegraphics[width=8cm]{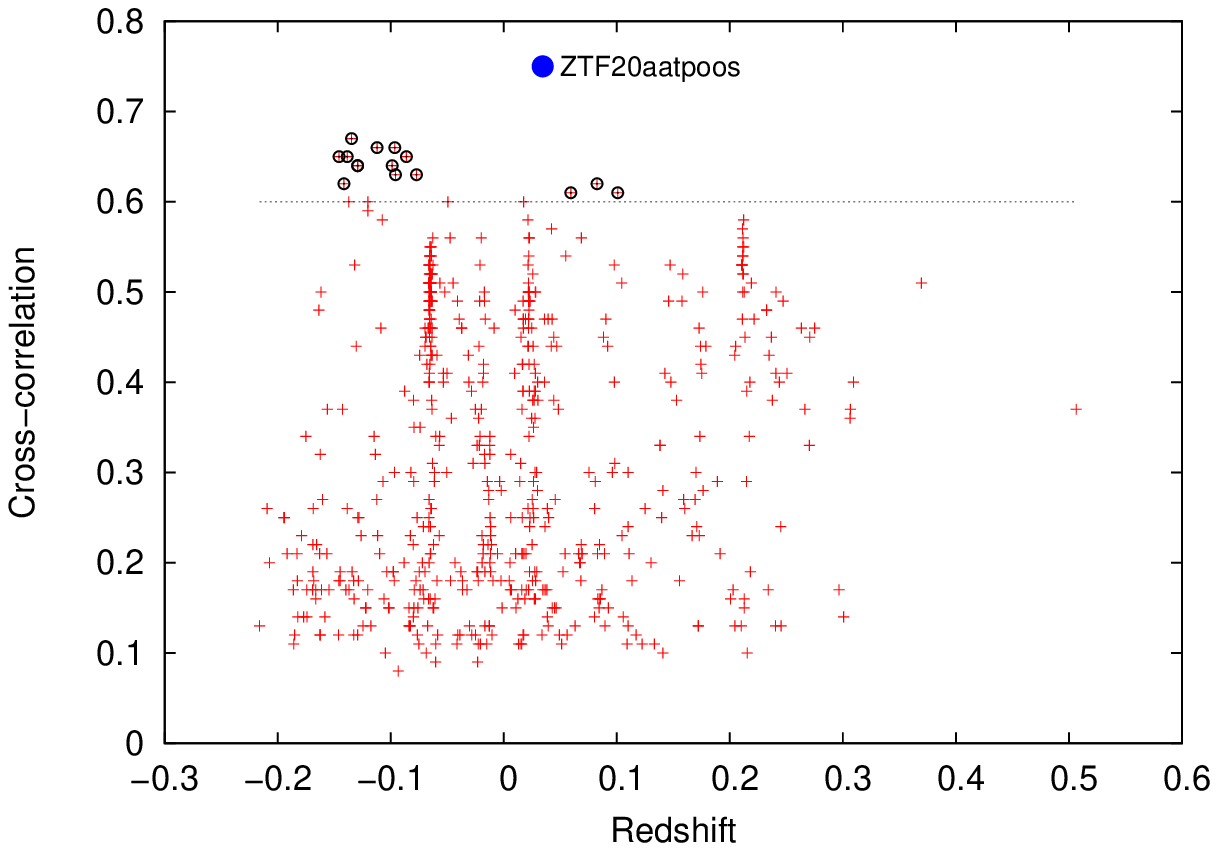}
    \caption{The maximum cross-correlation coefficient plotted against spectrum ID (left panel) and redshift (right panel) for the HETDEX sample. The dotted horizontal line marks the chosen threshold value for SN candidates (see text).}
    \label{fig:sn_ccf}
\end{figure*}

In the left panel of Figure~\ref{fig:sn_ccf}, the peak height of the \textit{CCF} is shown against the object index of the HETDEX spectra. It is seen that for the majority of objects, the \textit{CCF} is below 0.6, which we set as an initial threshold:  each spectrum exceeding this threshold was inspected visually to check whether the similarity with the SN template was real. 

In the right panel of Figure~\ref{fig:sn_ccf} the same quantity is plotted against the redshift provided by {\tt fxcor}. This diagram is useful because objects with negative redshifts can  immediately be rejected from the list of SN candidates, even though they may reach higher \textit{CCF} peaks. As a first approximation, any object above the 0.6 threshold level \textit{and} having positive redshift was considered a real candidate.

In both diagrams of Figure~\ref{fig:sn_ccf}, the object that has the highest \textit{CCF} peak (by far) is ZTF20aatpoos; this source was also identified as a good SN candidate in the previous sections. All other candidates having $z > 0$ and $\mathit{CCF} > 0.6$ turned out to be broad-line AGNs with strong emission lines, thus showing that their spectral overlap with SN templates is artificial. 

After the visual inspection of all HETDEX spectra of ``real'' (i.e., not {\tt BOGUS}, see Section~\ref{sec:chi2template}) sources, we also identified ZTF19abdkelq as a potential SN candidate. This spectrum was classified as {\tt low} by the $\chi^2$ template matching algorithm because of its low signal-to-noise parameter, but the \textit{CCF} method indicated some resemblance with the spectrum of Type Ia SN\null.  In this case, the peak of the \textit{CCF} is lower than the empirical threshold (0.6) due to the dominance of observational noise. 

More details on these two SN candidates, ZTF20aatpoos and ZTF19abdkelq, are presented in the next subsection.

\subsection{Supernovae}\label{sec:sne}

\begin{figure*}
    \centering
    \includegraphics[width=8cm]{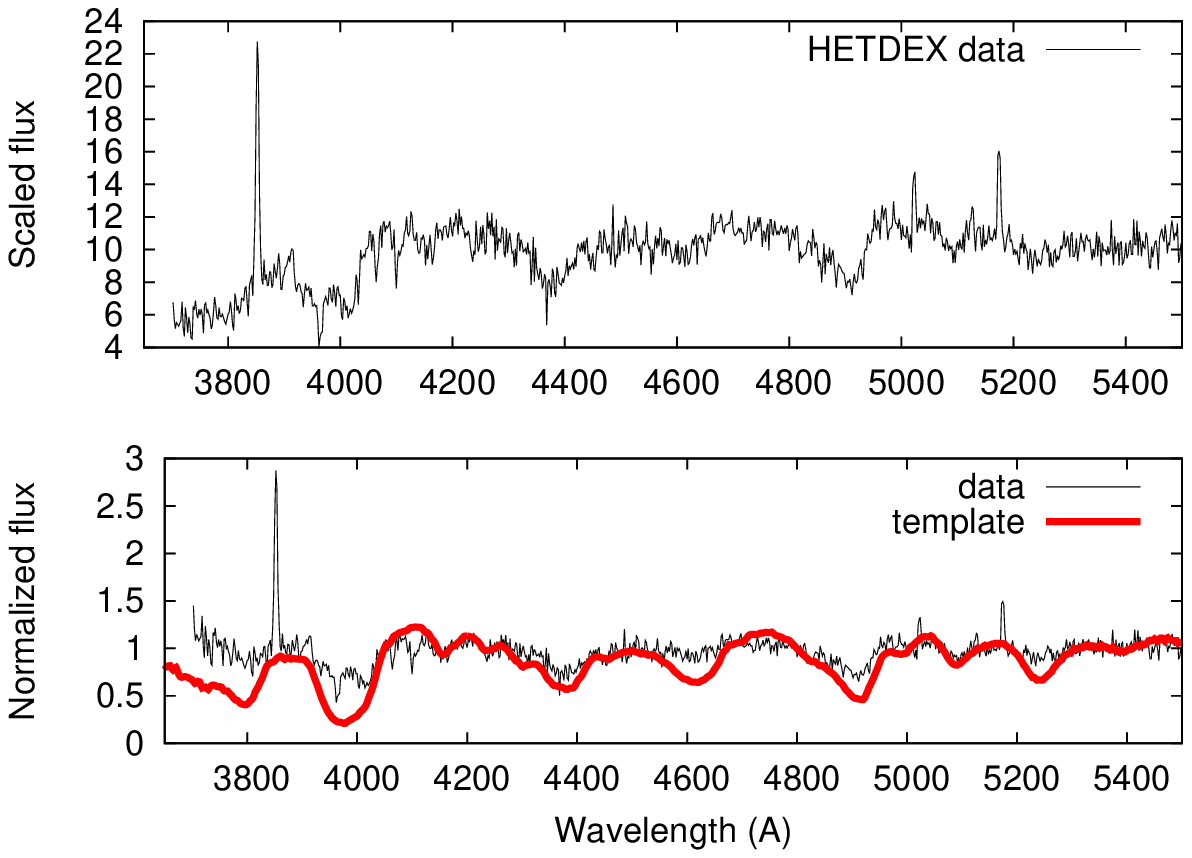}
    \includegraphics[width=8cm,height=6.2cm]{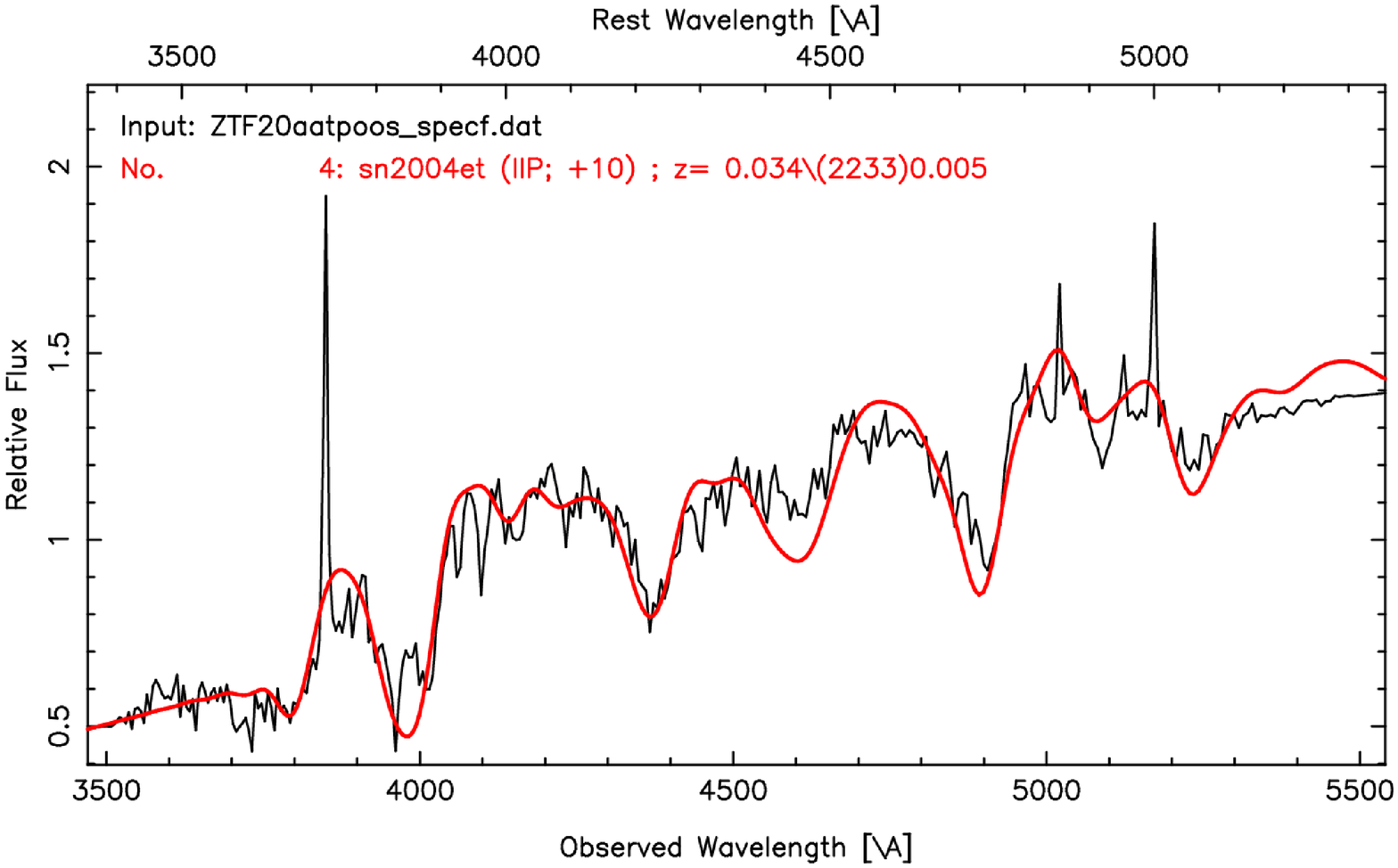}
    \caption{The spectrum of ZTF20aatpoos, a Type IIP SN at $\sim 10$ days after explosion (plotted as a black curve), compared to a Type IIP SN template (red curve). The left panel shows the result from our cross-correlation analysis (see Section~\ref{sec:disc-cc}), while the right panel displays the output from {\tt SNID}.}
    \label{fig:sn1}
\end{figure*}

The strongest candidate for being a SN in our HETDEX sample is ZTF20aatpoos. This object was originally classified as a SN by ALeRCE, and announced on the Transient Name Server (TNS) website\footnote{\tt https://www.wis-tns.org/} as AT~2020fiz on 2020-03-29 (see Table~\ref{tab:sne_snid}). HETDEX observed this target serendipitiously on 2020-04-15, $\sim 2$ weeks after discovery.

ZTF20aatpoos was identified as a SN candidate by all of our SN finding methods.  In Figure~\ref{fig:sn1} we plot its observed spectrum (black curve) together with the best-matching SN templates: our SN template (red curve) is shown in the bottom left panel, while a template found by {\tt SNID} is plotted with red on the right panel. It is seen that both \textit{CCF}-based codes found approximately the same SN template -- that for a relatively young (10 - 20 days after explosion) Type IIP SN.

\begin{figure}
    \centering
    \includegraphics[width=8cm]{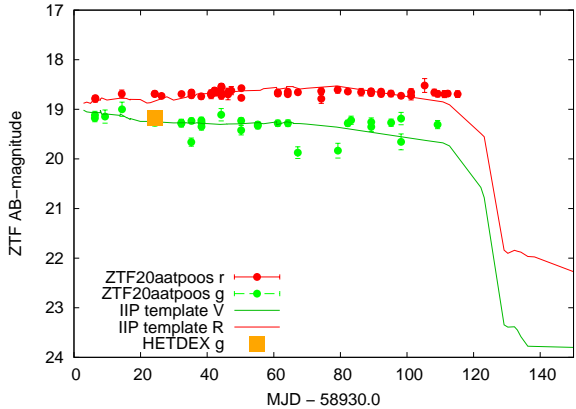}
    \caption{ZTF photometry on ZTF20aatpoos ($r$-band: red dots, $g$-band: green dots) compared to SN IIP template light curves, shifted vertically to match the observations. $g$-band  synthetic photometry from the HETDEX spectrum is plotted with an orange square. }
    \label{fig:phot_sn1}
\end{figure}

ZTF20aatpoos was followed up by ZTF for 109 days after discovery, and its light curve (plotted in Figure~\ref{fig:phot_sn1}) fully confirms that it shows the $\sim 100$ day-long plateau, characteristic of Type II-P SNe. Template light curves from the data of the Type IIP SN~2005cs \citep{pastor09} in $V$- and $R$-bands are shown for comparison. 

We also computed a synthetic $g$-band AB-magnitude for ZTF20aatpoos directly from the HETDEX spectrum. The value, $g = 19.180 \pm 0.002$ mag, is plotted as an orange square in Figure~\ref{fig:phot_sn1}. It is seen that it agrees very well with the $g$-band ZTF photometry taken at similar epochs. 

\begin{figure*}
    \centering
    \includegraphics[width=8cm]{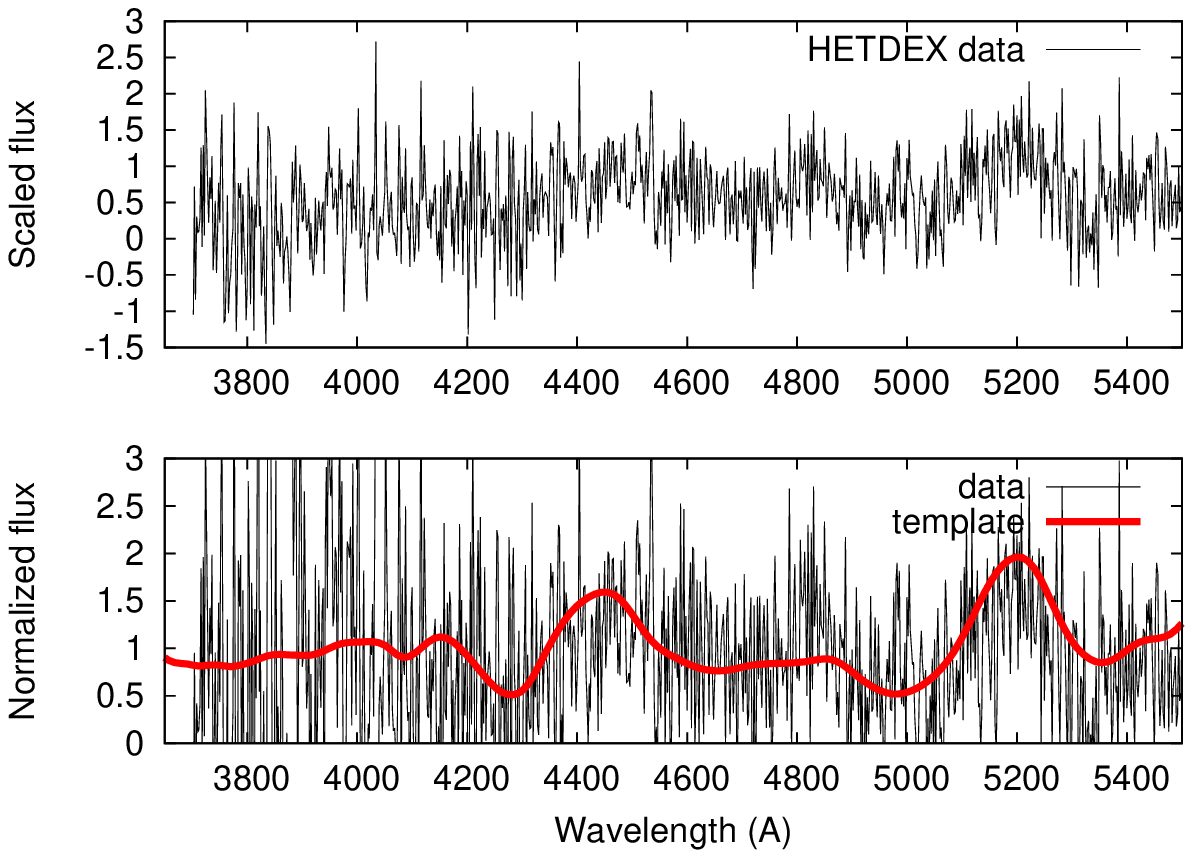}
    \includegraphics[width=8cm,height=6.2cm]{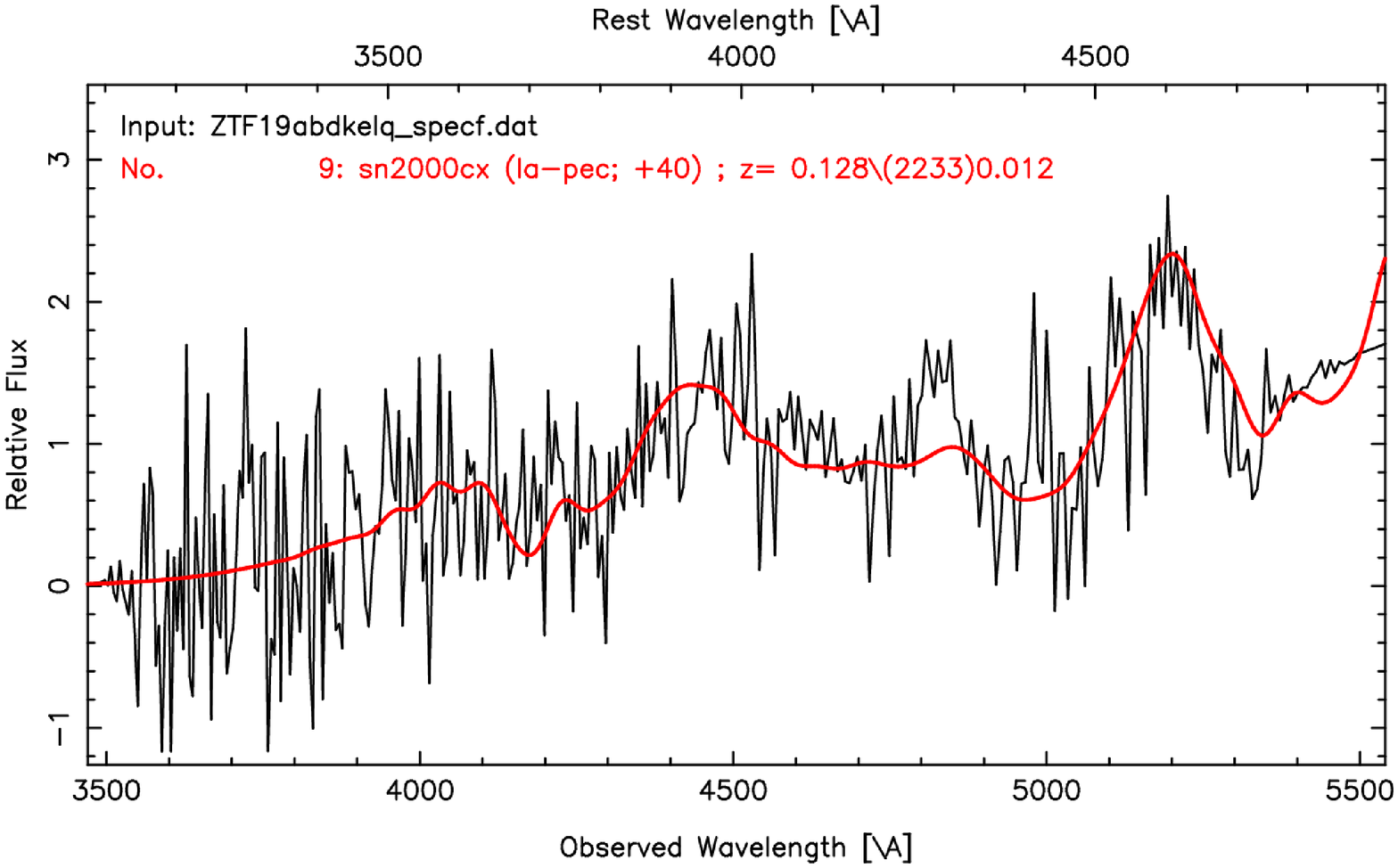}
    \caption{The same as Figure~\ref{fig:sn1} but for the Type Ia SN ZTF19abdkelq, observed at $\sim 40$ days after maximum light.}
    \label{fig:sn2}
\end{figure*}

\begin{figure}
    \centering
    \includegraphics[width=8cm]{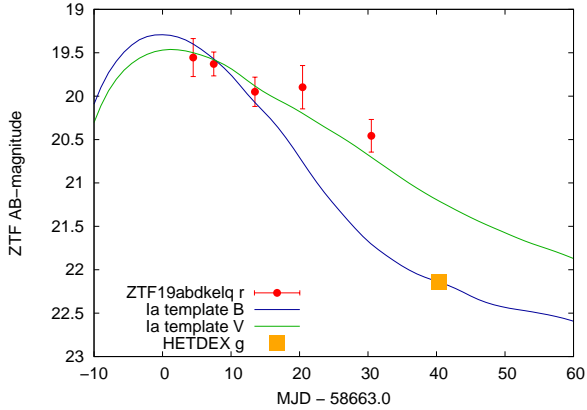}
    \caption{$r$-band ZTF photometry for ZTF19abdkelq, plotted together with Ia template light curves in rest-frame $B$ and $V$-bands. The $g$-band synthetic photometry from the HETDEX spectrum is shown by the orange square. }
    \label{fig:phot_sn2}
\end{figure}

\begin{table*}[]
    \centering
    \caption{Parameters of the SNe found in HETDEX}
    \begin{tabular}{lccccccc}
    \hline
    \hline
    ZTF name & IAU name & R.A.(deg) & Decl.(deg) & Date & {\tt SNID} type & {\tt SNID} $z$ & {\tt SNID} phase (day) \\ 
    \hline
    ZTF19abdkelq & -- & 23.086249 & 0.658040 & 2019-08-08 & Ia & 0.125 $\pm0.011$ & 40.4 $\pm$20.1 \\
    ZTF20aatpoos & AT 2020fiz & 186.512456 & 55.702264 & 2020-04-15 & IIP & 0.026 $\pm$0.013 & 10.7 $\pm$6.3 \\
    \hline
    \end{tabular}
    \label{tab:sne_snid}
\end{table*}

Our other SN candidate, ZTF19abdkelq, was originally classified as an asteroid ({\tt AST}) by ALeRCE, but ZTF also recorded $r$-band magnitudes at 4 later epochs, with the last ZTF detection occurring $\sim 30$ days after discovery. Thus, it is unlikely that this source is a moving object. Since it was not flagged as a transient, there is no available TNS record for the object.

The HETDEX spectrum of ZTF19abdkelq is plotted in Figure~\ref{fig:sn2}. In the left panel, we show the continuum-normalized spectrum (plotted with red) together with the best-matching Hsiao SN Ia template (black). Even though the observed spectrum is noisy, the pseudo-emission feature around 5200\,\AA\ is a good match to the template spectrum. The resemblance is strengthened by the analysis with {\tt SNID} (right panel), which suggests SN~2000cx, a peculiar Type Ia SN, as the best-matching template. Both our \textit{CCF} code and {\tt SNID} provided a consistent redshift for ZTF19abdkelq of $z \sim 0.12$. 

In Figure~\ref{fig:phot_sn2}, the light curve of ZTF19abdkelq is plotted from ZTF $r$-band photometry. Green and blue curves show the expected brightness decline of a Type Ia SN, as inferred from rest-frame $V$- and $B$-band template light curves from MLCS2k2 \citep{jha07}. At $z \sim 0.12$, the rest-frame $V$-band template light curve represents the observer-frame $r$-band data well. 
The result of $g$-band synthetic photometry from the HETDEX spectrum ($g = 22.144 \pm 0.098$ mag) is also shown as an orange square. It is seen that both the ZTF and the HETDEX photometry is entirely consistent with the proposed Type Ia SN classification of ZTF19abdkelq. 

The parameters for the two SNe found in HETDEX are summarized in Table~\ref{tab:sne_snid}. 

\subsection{Active Galactic Nuclei}
\label{sec:agn}

\begin{figure}
    \centering
    \includegraphics[width=8cm]{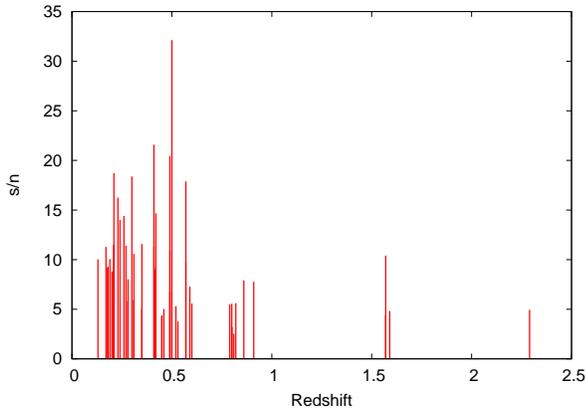}
    \caption{The {\tt s/n} parameter of the AGN spectra as a function of redshift.}
    \label{fig:agn_z}
\end{figure}

\begin{figure*}
    \centering
    \includegraphics[width=16cm]{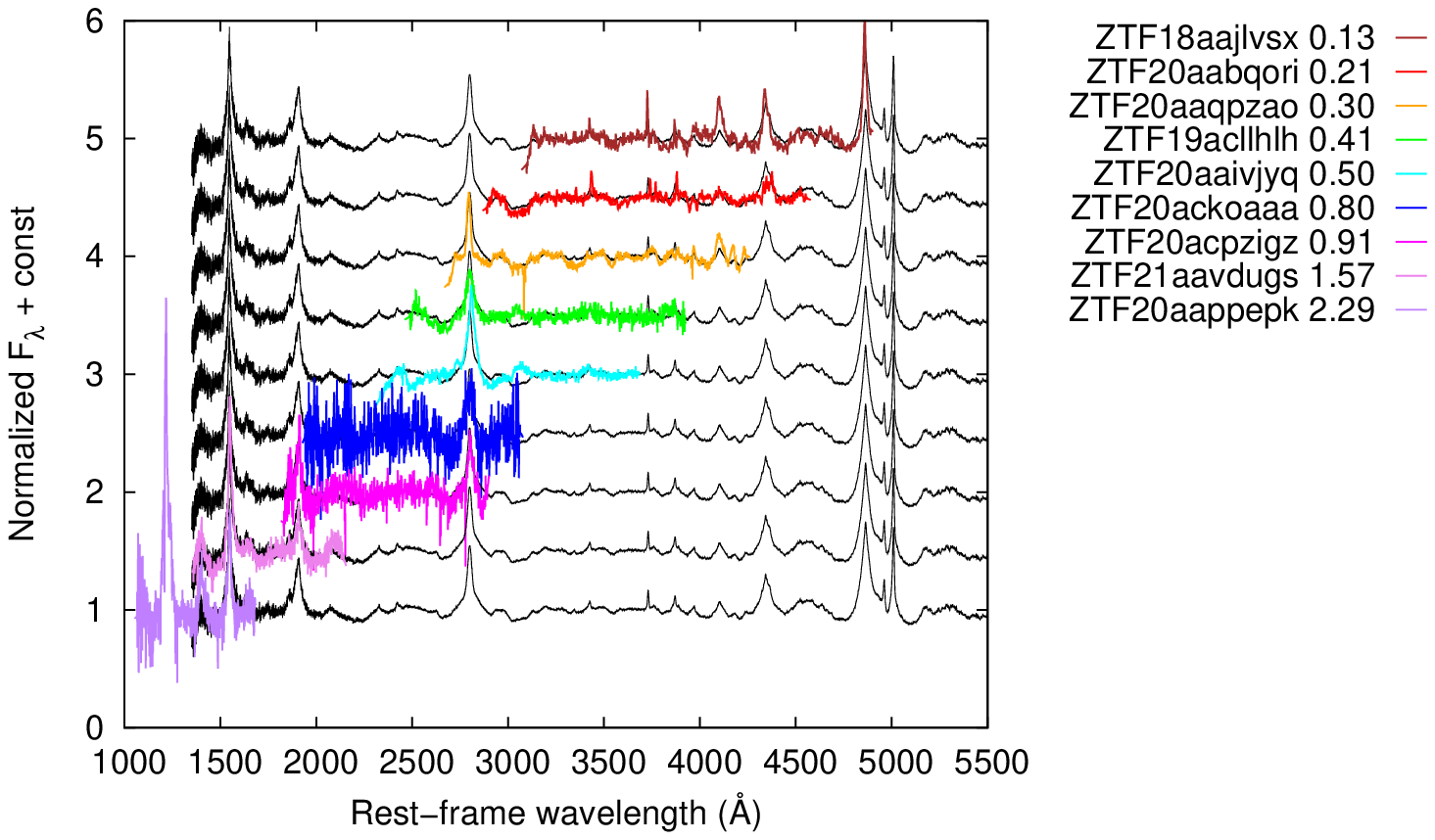}
    \caption{Observed spectra of a subsample of AGNs compared to the rest-frame SDSS QSO template.}
    \label{fig:agn_sp}
\end{figure*}

Members of the AGN group were identified via the $\chi^2$-minimization method (see Section~\ref{sec:chi2template}), although some were also found by cross-correlating the spectra with supernova templates (Section~\ref{sec:sntemplate}). We identified 52 spectra (Table~\ref{tab:stat}) that had a galaxy spectrum as their best-matching template, but also showed broader emission features than those of the star-forming galaxy templates. After visual inspection, 3 of them turned out to be saturated stars, which were then removed from the sample. The remaining 49 were analyzed further. 

Redshifts were derived by fitting the ``high luminosity QSO'' template spectrum taken from the SDSS site\footnote{\tt https://http://classic.sdss.org/dr5/algorithms/spectemplates/} to the AGN spectra. The fitting parameters for the AGN sample are collected in Table~\ref{tab:agn_z} in the Appendix.

Figure~\ref{fig:agn_z} shows the {\tt s/n} parameter of each spectrum (see Section~\ref{sec:chi2template}) against redshift. It is seen that most of the ZTF transients that turned out to be AGNs are at low redshifts, i.e., at $z \lesssim 0.5$. As a comparison, the full HETDEX AGN sample \citep{liu22} extends to at least $z \sim 3.5$ and has a median redshift of $z_{\rm med} \sim 2.1$; in other words, the AGNs identified here are more local than the majority of the HETDEX AGNs. This is most likely due to the sensitivity limit of the ZTF public stream, which is $m_{AB} \lesssim 20$ mag.  This limit is significantly brighter than the HETDEX detection threshold for AGNs \citep[$m_{AB} \sim 26$ mag;][]{liu22}. 

In Figure~\ref{fig:agn_sp}, spectra of some of the ZTF AGN sources are plotted against rest-frame wavelengths compared to the SDSS QSO template. One object, ZTF20aappepk ($z \sim 2.29$) has a HETDEX spectrum that extends beyond the blue edge of the SDSS QSO template, but the overlapping part shows good agreement with the template, just like the other, lower redshifted AGNs in this small sample.  

\subsection{Host galaxies of supernovae}
\label{sec:hosts}

\begin{figure}
    \centering
    \includegraphics[width=8cm]{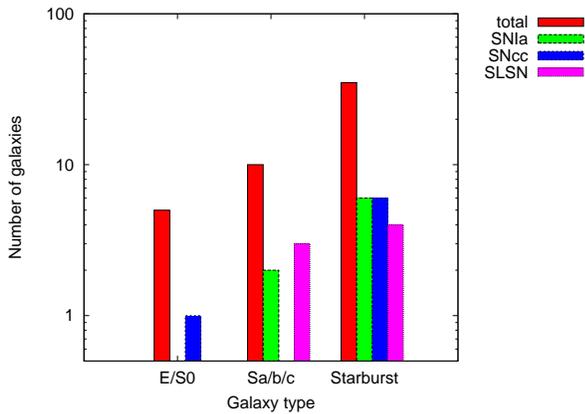}
    \caption{Distribution of host galaxy types for SNe with known light curve classification (green, blue and magenta bars) as well as that of the whole sample (red bars).}
    \label{fig:hostgx_hist}
\end{figure}

\begin{figure}
    \centering
    \includegraphics[width=8cm]{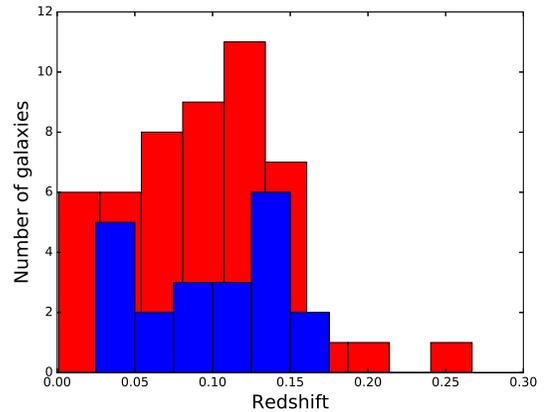}
    \caption{Histogram of host galaxy redshifts for the whole sample (red) and those having known ALeRCE light curve classification (blue).}
    \label{fig:hostgx_zdist}
\end{figure}

Information on the host galaxies of transients can be as important as the transients themselves. HETDEX presents a unique opportunity to gather spectra of transient host galaxies with a minimal observational cost: if a SN exploded within a HETDEX field, then it should be possible to extract the host galaxy spectrum from the HETDEX data products. While such a dedicated study of transient host galaxies is beyond the scope of the present study, it may be a fruitful endeavor for future work.

Here we present galaxy types and redshifts for those HETDEX galaxies that were extracted {\it at the position} of ZTF transients that were classified as SNe by ALeRCE\null. We found 50 such objects according to Table~\ref{tab:stat}, of which 21 had a refined ALeRCE supernova classification based on their ZTF light curves. ALeRCE uses the following categories from its 
{\tt lc\_classification\_transient} pipeline: {\tt SNIa}, {\tt SNIbc}, {\tt SNII} and {\tt SLSN}\null. This subsample is listed in Table~\ref{tab:hostgx} in the Appendix, together with the host galaxy properties found by the $\chi^2$-classification method (Section~\ref{sec:chi2template}). 

The distribution of the host galaxy types for this subsample is shown in Figure~\ref{fig:hostgx_hist}. Here we added the numbers of {\tt SNIbc} SNe to that of the {\tt SNII} objects, and use the {\tt SNcc} category for those core-collapse SNe.
Note that the numbers for the {\tt SLSN} category are probably overestimated by ALeRCE\null.   
The same distribution for the whole sample (i.e., with those objects that do not have an ALeRCE light curve classification) is also plotted via the red bars. It is clear that even though the numbers are too low for any robust conclusions, most of the SNe occurred in starburst galaxies showing strong emission features. Since core collapse SNe originating from massive stars likely dominate our low-redshift sample, the distribution of host galaxy types as shown in Figure~\ref{fig:hostgx_hist} is consistent with expectations. 

Note that ZTF19abuxhqd, which was classified as a likely Type II SN by ALeRCE, occurred in an early-type (S0) galaxy. This would be peculiar, since early-type galaxies usually host Type Ia SNe that originate from older population of stars; however, a close inspection of the noise-dominated ZTF light curve of ZTF19abuxhqd provides no convincing evidence that the presence of a core collapse SN\null.  Specifically, the transient showed random-like brightness fluctuations for $\sim 400$ days, instead of any sign of a decline. 

The distribution of the redshifts for the whole sample, as well as for those SNe that have refined ALeRCE types 
is plotted in Figure~\ref{fig:hostgx_zdist}. This figure confirms that the ZTF SNe whose host galaxies could be measured by HETDEX are distributed in the $z< 0.3$ low-redshift space.

\subsection{The expected number of supernovae in HETDEX}
\label{sec:rates}

Even though HETDEX is not an ideal survey for finding extragalactic transients (see Section~\ref{sec:intro}), the number of supernovae that are expected to show up within the survey footprint may be of interest. Since the classification of new transients generally needs spectroscopy, the future HETDEX data releases could be useful for classifying targets that might otherwise escape attention. 

The expected number of supernovae within the survey area ($\Omega$) covered by observations  during the survey time ($T_S$) in redshift interval $\Delta z$ at redshift $z$ can be expressed as 
\begin{equation}
    N_{SN}(z, \Delta z) = \Omega T_s \int_{z - \Delta z}^z \frac{R_{SN}}{1+z} \frac{dV}{dz} dz,
\end{equation}
where $R_{SN}$ is the measured supernova volumetric rate (the number of SNe per Mpc$^3$ per year) and $dV/dz$ is the differential comoving volume. The factor $1+z$ accounts for cosmological time dilation. 

The HETDEX survey field consists primarily of two parts: a Spring field of 390 deg$^2$ and a Fall field of 150 deg$^2$. Due to gaps between the IFUs, the effective survey field of view is $\sim 94$ deg$^2$ \citep{gebhardt21}. The area covered by a single shot with VIRUS is $\sim 0.015$ deg$^2$. To estimate the number of SNe found during a year, we assume that, on average, $\sim 3$ such fields are taken on each night in a year for either the Spring or the Fall field. For the volumetric SN rates, we adopt $R_{Ia} = 2.5 \times 10^{-5} \cdot (1+z)^{1.5}$ SNe~Mpc$^{-3}$~yr$^{-1}$ for SNe Ia  \citep{hounsell18} and $R_{CC}(z=0.028) = 9.1 \times 10^{-5}$ SNe~Mpc$^{-3}$~yr$^{-1}$ for core-collapse SNe \citep{frohmaier21}. The latter is scaled with the cosmic star formation rate function by \citet{hopkins06} for different redshifts.
Thus, using $\Omega = 3 \cdot 0.015 = 0.045$ deg$^2$, $T_s = 1$ year and the volumetric SN rates given above, we get $N_{Ia} \sim 0.24$ SN Ia in the $0 < z < 0.3$ redshift range. Since core collapse SNe are fainter by $\lesssim 1$ magnitude than Type Ia SNe, they are accessible by HETDEX only in the $0 < z < 0.2$ redshift range (assuming $\sim 22$ mag as the detection limit of continuum sources in HETDEX). Within this volume we get $N_{CC} \sim 0.38$ core collapse SN per year. 
This optimistic calculation does not take into account detection efficiencies (signal-to-noise, for example), thus, it is likely an overestimate, but the final numbers are consistent with our results (one SN of both types within $\sim5$ years of HETDEX operation). Note that between 2017 and 2022 the HETDEX field-of-view was not completely filled up by IFUs, which decreased our chances to find transients. Starting from 2022, having the field-of-view fully populated by 78 IFUs, the probability of discoveries are expected to increase. 
Overall, we conclude that we can expect to find a few more ($N \lesssim 5$) supernovae during the upcoming years of HETDEX operation.     

\section{Conclusions} \label{sec:conc}

The conclusions of the present study are summarized as follows:
\begin{itemize}
    \item{We identified the counterparts of 583 transient objects in the HETDEX spectral database (but not necessarily at the same epoch) out of 4845 transient sources found by ALeRCe in the ZTF public data stream. In total, 636 spectra covering the 3500 - 5500\,\AA\ wavelength range were extracted from the HETDEX observing archive and analyzed.}
    \item{We applied the t-distributed stochastic neighbor embedding (t-SNE) machine-learning method to explore the dataset and identify potentially interesting sources. Our primary intention was to identify supernovae, which succeeded in 2 cases (see below).}
    \item{All ZTF sources having HETDEX spectra were classified into {\tt star}, {\tt galaxy}, {\tt agn}, and {\tt low} categories based on matching with template spectra (or, in the last case, having low signal-to-noise). For most objects, these categories agreed very well with their ``Stamp-classification'' made by the ALeRCe pipeline for their ZTF counterparts.}
    \item{We attempted to find supernovae among the HETDEX spectra in three different ways: visual inspection of the t-SNE 2D output, looking for outliers in the $\chi^2$ template matching classification, and cross-correlation with SN templates. ZTF20aatpoos was flagged by all three methods as a potential SN, which was successfully confirmed as a Type IIP event using {\tt SNID}\null. We also identified another SN candidate, ZTF19abdkelq, among the low S/N sources, and {\tt SNID} confirmed it to be a Type Ia SN taken $\sim 40$ days after maximum light.}
    \item{As a by-product, we identified 49 AGN sources, most of them at low ($z \lesssim 0.5$) redshifts;  38 of these were also classified as AGN by ALeRCe.}
    \item{We obtained host galaxy spectra for 50 ZTF transients classified as supernovae by ALeRCE\null. Most turned out to be starburst galaxies, which is consistent with the expectation that core-collapse events dominate the supernova sample in the low-redshift ($z \lesssim 0.3$) Universe. }
\end{itemize}

We expect to find a few ($\lesssim 5$)  more supernovae during the upcoming years of the HETDEX survey.

\begin{acknowledgments}

HETDEX is led by the University of Texas at Austin
McDonald Observatory and Department of Astronomy with participation from the Ludwig-Maximilians Universitat Munchen, Max-Planck-Institut fur Extraterrestrische Physik (MPE), Leibniz-Institut fur Astrophysik Potsdam (AIP), Texas A\&M University, The Pennsylvania State University, Institut fur Astrophysik
Gottingen, The University of Oxford, Max-Planck Institut fur Astrophysik (MPA), The University of
Tokyo, and Missouri University of Science and Technology. In addition to Institutional support, HETDEX is
funded by the National Science Foundation (grant AST0926815), the State of Texas, the US Air Force (AFRL FA9451-04-2-0355), and generous support from private
individuals and foundations.

The Hobby-Eberly Telescope (HET) is a joint project
of the University of Texas at Austin, the Pennsylvania State University, Ludwig-Maximilians-Universitat
Munchen, and Georg-August-Universitat Gottingen.

The HET is named in honor of its principal benefactors, William P. Hobby and Robert E. Eberly.

VIRUS is a joint project of the University of Texas at Austin, Leibniz-Institut f{\" u}r Astrophysik Potsdam (AIP), Texas A\&M University (TAMU), Max-Planck-Institut f{\" u}r Extraterrestriche-Physik (MPE), Ludwig-Maximilians-Universit{\" a}t M{\" u}nchen, Pennsylvania State University, Institut f{\" u}r Astrophysik G{\" o}ttingen, University of Oxford, and the Max-Planck-Institut f{\" u}r Astrophysik (MPA).

The authors acknowledge the Texas Advanced Computing Center (TACC) at The University of Texas at
Austin for providing high performance computing, visualization, and storage resources that have contributed
to the research results reported within this paper. URL:
http://www.tacc.utexas.edu

The University of Texas at Austin sits on indigenous land. The Tonkawa lived in central Texas and the Comanche and Apache moved through this area. The Davis Mountains that host McDonald Observatory were originally husbanded by Lipan Apache, Warm Springs Apache, Mescalero Apache, Comanche and various tribes of the Jumanos. We acknowledge and pay our respects to all the Indigenous Peoples and communities who are or have been  a part of these lands and territories in Texas. 

JV, BPT and JCW are supported by NSF grant AST1813825. 

ZJ was supported by a PRODEX Experiment Agreement No. 4000137122 between the ELTE E\"otv\"os Lor\'and University and the European Space Agency (ESA-D/SCI-LE-2021-0025).

The Institute for Gravitation and the Cosmos is supported by the Eberly College of Science and the Office of the Senior Vice President for Research at the Pennsylvania State University.

\end{acknowledgments}

\facilities{Hobby-Eberly Telescope (HET); Zwicky Transient Facility (ZTF), Texas Advanced Computing Center (TACC)}

\software{
{\tt ALeRCE: http://alerce.science/}; \\
{\tt IRAF: https://iraf-community.github.io/} \citep{iraf86, iraf93}; \\
{\tt Astropy: http://www.astropy.org} \citep{astropy13, astropy18}; \\
{\tt numpy: https://numpy.org} \citep{numpy20}; \\
{\tt scipy: https://scipy.org } \citep{scipy20}; \\
{\tt matplotlib: https://matplotlib.org} \citep{matplotlib07}; \\
{\tt scikit-learn: https://scikit-learn.org/stable/} \citep{scikitlearn11}; \\
{\tt pandas: https://pandas.pydata.org/} \citep{pandas10}; \\
{\tt SNID: https://people.lam.fr/\\
blondin.stephane/software/snid/index.html} \citep{snid}.
}

\section{Appendix}

\setcounter{table}{0}
\renewcommand{\thetable}{A\arabic{table}}

\setcounter{figure}{0}
\renewcommand{\thefigure}{A\arabic{figure}}

Table~\ref{tab:chi2class} lists the parameters of the ZTF transients classified via $\chi^2$-minimization (Section~\ref{sec:chi2template}). Only a few objects are given here to illustrate the data structure. The full table in electronic form can be accessed on GitHub\footnote{\tt https://github.com/jozsefvinko/Transients-in-HETDEX}.

\begin{table*}[!ht]
\caption{ZTF transients classified via $\chi^2$-minimization }
\begin{center}
\scriptsize
\begin{tabular}{lrrcccccccccc}
\hline
\hline
Name & R.A.(deg) & Dec.(deg) & MJD$_{\rm start}$ & MJD$_{\rm stop}$ & MJD$_{\rm DEX}$ & AL type & AL prob. & Sp.type & S/N & $\chi^2$ & z & Class. \\ 
\hline
ZTF17aaclhdu & 33.36269135 & -0.87615907 & 58781.3 & 59079.5 & 58862 & AGN & 0.46 & G0V & 16.44 & 0.9381 & 0 & star \\
ZTF18aagstka & 226.93757719 & 51.45284761 & 58255.3 & 59641.4 & 58660 & SN & 0.43 & SB4 & 3.12 & 0.2609 & 0.045 & galaxy \\
ZTF18aaguppa & 230.02485526 & 51.14954140 & 59067.2 & 59094.2 & 58173 & -- & -- & M2V & 0.96 & 1.3199 & 0 & low \\
ZTF18aaiwfgo & 197.17922553 & 55.52398578 & 59291.3 & 59316.3 & 58958 & AGN & 0.65 & Sa & 8.02 & 1.9890 & 0.079 & agn \\
$\vdots$ & $\vdots$ & $\vdots$ & $\vdots$ & $\vdots$ & $\vdots$ & $\vdots$ & $\vdots$ & $\vdots$ & $\vdots$ & $\vdots$ & $\vdots$ & $\vdots$ \\
\hline
\end{tabular}
\label{tab:chi2class}
\end{center}
\end{table*}

\begin{table*}[!ht]
\caption{Parameters of the AGN sample}
\centering
\begin{tabular}{lrrc|lrrc}
\hline
\hline
ZTF name & S/N & $\chi^2$ & $z$ & ZTF name & S/N & $\chi^2$ & $z$ \\
\hline
ZTF18aaiwfgo & 8.01 & 1.2152 & 0.28 & ZTF20aarsdui & 4.37 & 8.0946 & 1.57 \\
ZTF18aajlvsx & 10.02 & 0.6340 & 0.13 & ZTF20aasjgqr & 5.47 & 0.9711 & 0.79 \\
ZTF18aakqshm & 16.24 & 1.2993 & 0.23 & ZTF20aaxalny & 11.57 & 0.9945 & 0.35 \\
ZTF18aaktpmg & 9.21 & 11.2259 & 0.42 & ZTF20abakioq & 4.31 & 2.0508 & 0.45 \\
ZTF18aaqkxzg & 10.05 & 0.4965 & 0.19 & ZTF20abbfimy & 4.82 & 2.3148 & 1.59 \\
ZTF18aaxqtbj & 13.99 & 1.2383 & 0.24 & ZTF20abcamux & 9.19 & 0.4522 & 0.18 \\
ZTF18aceyycp & 9.22 & 0.6280 & 0.21 & ZTF20abcjsvg & 2.53 & 0.3731 & 0.81 \\
ZTF19aalbcut & 14.66 & 4.4769 & 0.42 & ZTF20abrigbg & 5.05 & 0.6352 & 0.35 \\
ZTF19aarioyj & 11.27 & 1.3345 & 0.17 & ZTF20abwfqoj & 9.74 & 5.6799 & 0.57 \\
ZTF19aasbotx & 11.42 & 1.2587 & 0.27 & ZTF20abxpvdk & 5.57 & 1.7046 & 0.60 \\
ZTF19aasccum & 5.30 & 2.6832 & 0.52 & ZTF20abxrxbh & 17.89 & 3.9041 & 0.57 \\
ZTF19acllhlh & 21.58 & 1.6776 & 0.41 & ZTF20ackoaaa & 5.54 & 0.9043 & 0.80 \\
ZTF20aabqori & 18.72 & 1.0551 & 0.21 & ZTF20acpzigz & 7.76 & 0.7796 & 0.91 \\
ZTF20aadbtju & 10.57 & 1.6565 & 0.31 & ZTF20acxtdub & 11.50 & 1.5784 & 0.21 \\
ZTF20aahgivv & 8.96 & 1.5503 & 0.42 & ZTF21aandbyi & 4.36 & 2.0195 & 0.45 \\
ZTF20aaicqat & 5.61 & 1.0932 & 0.82 & ZTF21aaplbpm & 5.02 & 3.9342 & 0.46 \\
ZTF20aaivjyq & 32.11 & 6.4454 & 0.50 & ZTF21aavdhbj & 3.18 & 1.8110 & 0.80 \\
ZTF20aajbuhw & 5.79 & 0.9001 & 0.28 & ZTF21aavdugs & 10.41 & 1.1423 & 1.57 \\
ZTF20aajcmsc & 9.30 & 0.4392 & 0.18 & ZTF21aayehok & 6.74 & 3.4700 & 0.49 \\
ZTF20aaliybg & 14.40 & 1.7369 & 0.26 & ZTF21aazcpmg & 3.77 & 2.0711 & 0.53 \\
ZTF20aankago & 7.90 & 0.9189 & 0.86 & ZTF21aazrcye & 5.93 & 0.8802 & 0.31 \\
ZTF20aaoyjbk & 10.92 & 1.9170 & 0.49 & ZTF21aazxsci & 8.80 & 1.2792 & 0.20 \\
ZTF20aaoykbt & 11.35 & 1.6181 & 0.41 & ZTF21abatxlh & 7.27 & 3.7249 & 0.59 \\
ZTF20aappepk & 4.93 & 2.3970 & 2.29 & ZTF21abiimfm & 20.44 & 2.4730 & 0.49 \\
ZTF20aaqpzao & 18.37 & 1.2529 & 0.30 & & & & \\
\hline
\end{tabular}
\label{tab:agn_z}
\end{table*}

\begin{table*}[!ht]
\caption{Parameters of the host galaxies having ALeRCE SN light curve classification}
\centering
\begin{tabular}{lcccccc}
\hline
\hline
Name & SN type & SN type prob. & Host type & S/N & $\chi^2$ & z \\
\hline
ZTF18aagstka & SNII & 0.408 & SB4 & 3.12 & 0.2609 & 0.045 \\
ZTF18aamsecj & SNII & 0.296 & SB6 & 9.27 & 2.3834 & 0.080 \\
ZTF18aaovsji & SLSN & 0.320 & SB1 & 6.96 & 6.8908 & 0.131 \\
ZTF18abkiqna & SNIa & 0.330 & SB4 & 5.10 & 4.3505 & 0.130 \\
ZTF18abuatfp & SNII & 0.342 & SB6 & 11.81 & 5.3940 & 0.136 \\
ZTF18acwtrfe & SNIbc & 0.336 & SB3 & 4.64 & 1.0119 & 0.049 \\
ZTF18acwyxnp & SLSN & 0.364 & Sa & 12.32 & 3.3459 & 0.130 \\
ZTF19aangier & SNIa & 0.472 & Sb & 14.44 & 2.5332 & 0.060 \\
ZTF19abuxhqd & SNII & 0.346 & S0 & 13.96 & 3.1949 & 0.080 \\
ZTF19aceckmm & SNII & 0.462 & SB6 & 7.45 & 2.5238 & 0.032 \\
ZTF19acfxbki & SNIa & 0.596 & Sa & 21.03 & 2.3064 & 0.058 \\
ZTF20aafcheh & SLSN & 0.418 & Sa & 12.44 & 3.3433 & 0.130 \\
ZTF20aagnbdf & SLSN & 0.438 & SB3 & 5.86 & 3.0425 & 0.089 \\
ZTF20aahhvci & SNIa & 0.382 & SB3 & 4.32 & 2.0259 & 0.104 \\
ZTF20aaolglj & SLSN & 0.360 & SB3 & 9.39 & 1.6336 & 0.029 \\
ZTF20aausyrp & SNIa & 0.296 & SB6 & 5.10 & 3.3010 & 0.026 \\
ZTF20aayxapt & SLSN & 0.376 & SB1 & 8.35 & 4.8623 & 0.145 \\
ZTF20abeiqzy & SLSN & 0.344 & Sa & 9.53 & 2.4443 & 0.171 \\
ZTF21aaoblpm & SNIa & 0.384 & SB3 & 12.35 & 13.0117 & 0.117 \\
ZTF21aaplbom & SNIbc & 0.360 & SB3 & 11.24 & 6.0530 & 0.121 \\
ZTF21abfjcap & SNIa & 0.388 & SB6 & 12.76 & 3.6394 & 0.151 \\
\hline
\end{tabular}
\label{tab:hostgx}
\end{table*}

\end{document}